\documentclass[%
 reprint,
 amsmath,amssymb,
 aps, prl,
floatfix,
]{revtex4-1}

\usepackage{graphicx}
\usepackage{dcolumn}
\usepackage{bm}
\usepackage{xcolor}
\usepackage{soul}

\usepackage{subcaption}
\usepackage{placeins}
\usepackage{dblfloatfix}
\usepackage{upgreek}

\begin{document}


\preprint{APS/123-QED}

\title{Prevention of core particle depletion in stellarators by turbulence}
\author{H.\ Thienpondt\vspace{0.1cm}}%
\email{Hanne.Thienpondt@ciemat.es\vspace{0.2cm}}
\affiliation{%
	Laboratorio Nacional de Fusi\'on, CIEMAT, 28040 Madrid, Spain\\
}%
\author{J.\ M.\ Garc\'ia-Rega\~na}
\affiliation{%
	Laboratorio Nacional de Fusi\'on, CIEMAT, 28040 Madrid, Spain\\
}%
\author{I.\ Calvo}%
\affiliation{%
	Laboratorio Nacional de Fusi\'on, CIEMAT, 28040 Madrid, Spain\\
}%
\author{J.\ A.\ Alonso}%
\affiliation{%
	Laboratorio Nacional de Fusi\'on, CIEMAT, 28040 Madrid, Spain\\
}%
\author{J.\ L.\ Velasco}%
\affiliation{%
	Laboratorio Nacional de Fusi\'on, CIEMAT, 28040 Madrid, Spain\\
}%
\author{A.\ Gonz\'alez-Jerez}%
\affiliation{%
	Laboratorio Nacional de Fusi\'on, CIEMAT, 28040 Madrid, Spain\\
}%
\author{\vspace{-0.5cm}M. Barnes\vspace{0.1cm}}%
\affiliation{%
	Rudolf Peierls Centre for Theoretical Physics, University of Oxford, Oxford OX1 3NP, United Kingdom\\
}%
\author{\vspace{-0.5cm}K. Brunner\vspace{0.1cm}}%
\affiliation{%
	Max-Planck-Institut f\"ur Plasmaphysik, 17491 Greifswald, Germany\\
}%
\author{O. Ford}%
\affiliation{%
	Max-Planck-Institut f\"ur Plasmaphysik, 17491 Greifswald, Germany\\
}%
\author{G. Fuchert}%
\affiliation{%
	Max-Planck-Institut f\"ur Plasmaphysik, 17491 Greifswald, Germany\\
}%
\author{J. Knauer}%
\affiliation{%
	Max-Planck-Institut f\"ur Plasmaphysik, 17491 Greifswald, Germany\\
}%
\author{E. Pasch}%
\affiliation{%
	Max-Planck-Institut f\"ur Plasmaphysik, 17491 Greifswald, Germany\\
}%
\author{L. Van\'o}%
\affiliation{%
	Max-Planck-Institut f\"ur Plasmaphysik, 17491 Greifswald, Germany\\
}%
\author{\vspace{-0.5cm}and the Wendelstein 7-X Team\vspace{0.3cm}}

\date{\today} 


\begin{abstract}
	In reactor-relevant plasmas, neoclassical transport drives an outward particle flux in the core of large stellarators and predicts strongly hollow density profiles. However, this theoretical prediction is contradicted by experiments. In particular, in Wendelstein 7-X, the first large optimized stellarator, flat or weakly peaked density profiles are generally measured, indicating that neoclassical theory is not sufficient and that an inward contribution to the particle flux is missing in the core. In this Research Letter, it is shown that the turbulent contribution to the particle flux can explain the difference between experimental measurements and neoclassical predictions. The results of this Research Letter also prove that theoretical and numerical tools are approaching the level of maturity needed for the prediction of equilibrium density profiles in stellarator plasmas, which is a fundamental requirement for the design of operation scenarios of present devices and future reactors.
\end{abstract}

\maketitle


\section{\label{sec:intro} Introduction}

The stellarator is the main alternative to the tokamak in the quest for achieving net energy from controlled nuclear fusion, due to its intrinsic steady-state operation and absence of current-driven instabilities. Nonetheless, neoclassical transport---associated with the inhomogeneity of the magnetic field and with particle collisions---has traditionally handicapped the performance of stellarators. For this reason, its minimization is one of the main targets in stellarator optimization. Wendelstein 7-X (W7-X) has demonstrated that large stellarators with optimized neoclassical transport can be designed \cite{Grieger_fst_1992} and built with high accuracy \cite{Pedersen_nature_2016} and that this has been fundamental for achieving triple product record values \cite{Beidler_nature_2021}.

However, even in optimized stellarators, neoclassical transport may play a major role in the plasma core \cite{Dinklage_nf_2013}. In this region, high temperatures are achieved, causing strong neoclassical particle and energy fluxes due to their unfavorable scaling with the temperature in low-collisionality plasma regimes. For example, in the $1/\nu$ regime, the neoclassical particle and energy fluxes scale as $\smash{T_s^{7/2}}$ and $\smash{T_s^{9/2}}$, respectively, with $T_s$ being the temperature of species $s$ (see, e.g., Ref.~\cite{Beidler_nature_2021}). In what follows, we only deal with plasmas consisting of singly charged ions and electrons, so that $s$ can take the values $s = i$ and $e$, respectively.

\begin{figure}
	\includegraphics{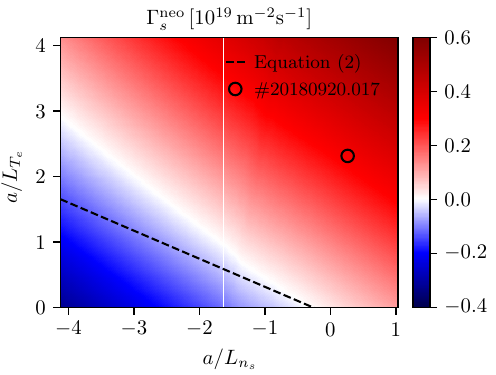}\vspace{-1mm}
	\caption{Neoclassical particle flux as a function of $a/L_{T_e}$ and $a/L_{n_s}$ for the standard W7-X configuration \cite{Geiger_2015}. Here, $r/a=0.25$, $n_e\!=\!n_i\!=6.3 \times 10^{19}\,$m$^{-3}$, $T_e=2.2\,$keV, $T_i=1.1\,$keV and $a/L_{T_i}=0.81$. The circle corresponds to the experimental discharge analyzed in the final part of this Letter. The dashed line shows the combination of $a/L_{T_e}$ and $a/L_{n_s}$ that gives $\Gamma^{\mathrm{neo}}_s=0$ in Eq.\,(\ref{EQ_DN}).}
	\vspace{-3mm}
	\label{fig:knosos_2d_scan}
\end{figure}

Stellarator plasmas with core particle transport dominated by neoclassical processes have been predicted to present problems of core particle depletion \cite{Maassberg_ppcf_1999, Beidler_ppcf_2018}. 
If the core particle source is negligible, as is the case in large stellarators with peripheral fueling, the steady-state particle balance equation for each of the species reads $\Gamma_s=\Gamma^{\mathrm{neo}}_s+\Gamma^{\mathrm{turb}}_s=0$. Here, $\Gamma_s$ is the flux-surface-averaged radial particle flux, and it is assumed to be the sum of a neoclassical contribution $\Gamma^{\mathrm{neo}}_s$ and a turbulent contribution $\Gamma^{\mathrm{turb}}_s$. Assuming $\Gamma^{\mathrm{turb}}_s=0$, the above steady-state conditions reduce to $\Gamma^{\mathrm{neo}}_i=\Gamma^{\mathrm{neo}}_e=0$. This relation, which additionally implies that no net radial current exists in the plasma, i.e., that the particle fluxes are ambipolar, imposes a constraint on the plasma profiles. The neoclassical particle flux can be written as 
\begin{equation}
\frac{{\Gamma_s^{\mathrm{neo}}}}{n_s}= - L_{11}^s\left(\frac{1}{n_s}\frac{dn_{s}}{dr}- \frac{Z_seE_r}{T_s}+\delta_{12}^s\frac{1}{T_s}\frac{dT_{s}}{dr}\right)\,,
\label{EQ_GNC}
\end{equation}
where $r\in[0,a]$ is a radial coordinate labeling magnetic surfaces, $a$ is the minor radius, $E_r$ is the radial electric field, $n_s$ is the density, and $Z_se$ is the electric charge ($Z_i = 1$ for ions and $Z_e = -1$ for electrons). Quasineutrality implies $n_i=n_e$, and its radial derivative implies $dn_i/dr = dn_e/dr$. Equation (\ref{EQ_GNC}) is valid at low collisionality and has been discussed in detail in Ref.~\cite{Beidler_nature_2021} and references therein. The neoclassical transport coefficients $L_{11}^s$ and $\delta_{12}^s$ generally depend on the magnetic configuration and plasma parameters, including $E_r$. However, if the electrons and main ions are assumed to be in the neoclassical $1/\nu$ and $\sqrt{\nu}$ collisionality regimes, respectively---the relevant regimes in the core of reactor-relevant stellarator plasmas---we can approximate $\delta_{12}^e=7/2$ and $\delta_{12} ^i=5/4$, and, after imposing $\Gamma^{\mathrm{neo}}_i=\Gamma^{\mathrm{neo}}_e=0$, arrive  at the constraint on the plasma profiles \cite{Beidler_ppcf_2018}. Namely,
\begin{equation}
\frac{1}{n_e}\frac{dn_{e}}{dr}=
-\frac{7/2}{T_e+T_i}\frac{dT_{e}}{dr} -\frac{5/4}{T_e+T_i}\frac{dT_{i}}{dr}\,.
\label{EQ_DN}
\end{equation}\vskip-0.0cm
\noindent Reactor-relevant plasmas are expected to display peaked temperature profiles, $dT_{s}/dr<0$, with $T_e\approx T_i$. Since $\delta_{12}^s>0$, this automatically leads to a hollow density profile $dn_{e}/dr>0$. Because $\delta_{12}^s>1$, (i.e., because the so-called thermodiffusion coefficient $L_{11}^s\delta_{12}^s$ is relatively large), the hollowness can become large enough to make the pressure gradient positive.

The above analytical argument illustrates the generality of the mechanism of core depletion by neoclassical transport. Of course, more accurate relations between the stationary density and temperature gradients can be obtained by means of numerical simulations. Figure \ref{fig:knosos_2d_scan} shows the neoclassical radial particle flux $\Gamma_s^{\mathrm{neo}}$, obtained with the neoclassical code \texttt{KNOSOS} \cite{Velasco_nf_2021}, over a wide range of values for the density and electron temperature gradients. In Fig.~\ref{fig:knosos_2d_scan} and in the remainder of this Research Letter, we define $a/L_{T_s}:=-a \, d\ln T_{s}/d r$ and $a/L_{n_s}:=-a\,d\ln n_{s}/d r$. Peaked (hollow) density profiles have $a/L_{n_s} >0$ ($a/L_{n_s} < 0$). For the rest of the parameters,
typical values from W7-X hydrogen plasmas sustained by Electron Cyclotron Resonance Heating (ECRH) are employed (see the caption of Fig.~\ref{fig:knosos_2d_scan} and Ref.~\cite{Beurskens_nf_2021}), and $E_r$ is set by ambipolarity of the neoclassical fluxes.
Outward fluxes ($\Gamma_s^\text{neo} > 0$) are obtained in most of the represented area, driven by $a/L_{T_e}$, indicating the importance of thermodiffusion. Indeed, the contour $\Gamma_s^{\mathrm{neo}}=0$ (white region) predicts a slightly hollow stationary density profile even in the presence of a practically flat electron temperature profile. For the larger values of $a/L_{T_e}$ obtained in W7-X, a distinctive hollow density profile should be measured if particle transport were described by neoclassical theory only.

Despite this robust neoclassical prediction, flat or weakly peaked density profiles have generally been measured in ECRH plasmas of W7-X (see,  e.g., Ref.~\cite{Wolf_nf_2017}). This disagreement between neoclassical theory and experiments may indicate that a significant inward contribution ($\Gamma_s < 0$) to the particle flux has not yet been identified. Under this hypothesis, this Research Letter addresses the particle transport problem in W7-X from the perspective of microturbulence and by means of a large number of nonlinear gyrokinetic simulations in stellarator geometry, carried out with the code \texttt{stella} \cite{Barnes_jcp_2019}. First, a parameter scan is performed, demonstrating that the turbulence resulting from the onset of the ion and the electron temperature gradients produces an inward pinch in broad regions of parameter space. Then, a specific W7-X discharge is studied, performing turbulence simulations throughout the plasma radius. The resulting profile of the turbulent particle flux is compared with the shortfall in the flux inferred from a careful particle balance analysis that includes estimates of the particle source and computations of the neoclassical particle flux. The sign of the missing contribution to the particle flux needed to sustain the experimental density profile agrees with that of the calculated turbulent particle flux, not only at the core but also at the edge of the plasma. In particular, the simulations consistently predict a sign change of the turbulent flux at intermediate radial positions, reflecting the fact that the neoclassical flux is too large at the core and too small at the edge of the plasma to explain the experimentally determined particle flux.


\begin{figure}[h]
	\centering
	\vspace{2mm}
	\includegraphics[trim={0mm 0mm 0mm 0mm},clip]{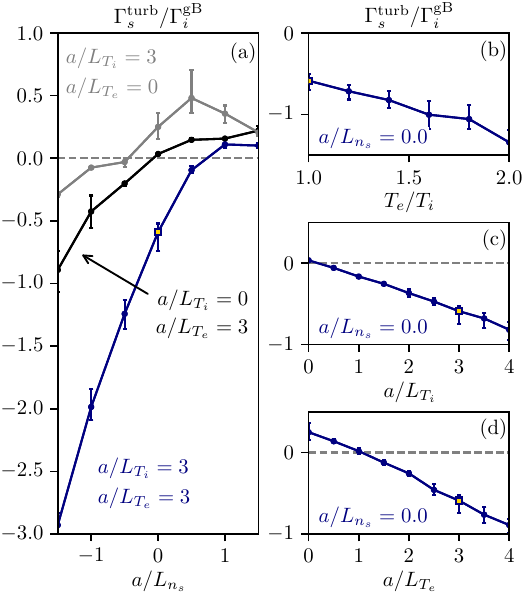} \vspace{-4mm}
	\caption{Turbulent radial particle flux in gyro-Bohm units  as a function of the (a) normalized density gradient, (b) electron-temperature-to-ion-temperature ratio, (c) normalized ion temperature gradient, and (d) normalized electron temperature gradient. The point with $T_e/T_i=1$, $a/L_{T_i} = a/L_{T_e} = 3$, and $a/L_{n_s}=0$ (golden square) is  shared by the four represented scans.} 
	\vspace{-1mm}
	\label{fig:scans}
\end{figure}

\section{Parametric dependence of the turbulent particle flux} 
\label{sec:parameterscans}

Let us write the radial turbulent particle flux as the sum of a diffusive and a convective contribution,
\begin{equation}
\frac{\Gamma_s^\text{turb}}{n_s} = -D \frac{1}{n_s}\frac{dn_s}{dr} + V\,,
\end{equation} 
with $D$ being the diffusion coefficient (which, in general, depends on the density gradient) and $V$ being the convection velocity. As $D$ is always positive, the diffusive term adds a contribution to the flux with opposite direction to that of the density gradient. In other words, the diffusive flux is inwardly or outwardly directed depending on whether the density gradient is hollow or peaked, respectively. On the other hand, $V$ can be either positive or negative, supporting the formation of hollow or peaked density profiles, respectively. More specifically, the sign of $V$ for a flat density profile ($a/L_{n_s}=0$) is of	particular interest, as it predicts whether a hollow (if $V>0$) or peaked (if $V<0$) density profile will develop. In order to gain insight into this matter, and prior to the comparison between numerical and W7-X experimental results, we provide a comprehensive numerical characterization of the turbulent particle flux, with emphasis on the case with a flat density profile, i.e., $a/L_{n_s}=0$.

The turbulent transport is modeled with the flux-tube $\delta\!f$ gyrokinetic code \texttt{stella} \cite{Barnes_jcp_2019}, which has been extensively benchmarked for W7-X geometry \cite{GonzalezJerez_jpp_2022} and applied to the study of turbulent impurity transport in this device~\cite{Regana_jpp_2021, Regana_nf_2021}. The gyrokinetic simulations are nonlinear, collisionless, electrostatic and account for kinetic ions and electrons. The calculations are performed at $r/a\!=\!0.25$ for the standard W7-X configuration. Due to the low magnetic shear, we employed generalized twist-and-shift boundary conditions \cite{Martin_ppcf_2018}. Further information on the resolution, boundary conditions, flux tube employed and other details of the gyrokinetic simulations presented in this Research Letter, can be found in Appendix A.

We carried out a parameter scan around the reference point in parameter space $\left\{T_e/T_i, a/L_{T_i}, a/L_{T_e}, a/L_{n_s}\right\}=\left\{1, 3, 3, 0\right\}$ in order to study the dependence of the turbulent particle flux on the following quantities: $a/L_{n_s}$ in Fig.~\ref{fig:scans}(a), the electron-temperature-to-ion-temperature ratio $T_e/T_i$ in Fig.~\ref{fig:scans}(b), the normalized ion temperature gradient $a/L_{T_i}$ in Fig.~\ref{fig:scans}(c), and the normalized electron temperature gradient $a/L_{T_e}$ in Fig.~\ref{fig:scans}(d). Such exhaustive scans cover both reactor-relevant conditions, where ions and electrons are expected to be thermalized and have comparable values of density and temperature gradients, and conditions corresponding to the first W7-X campaigns, where the temperature and its gradient are noticeably larger for electrons than for ions in the core. For the sake of clarity, the reference point in parameter space is represented with a golden square. 

\begin{figure}[b] 
	\vspace{-1mm}
	\includegraphics{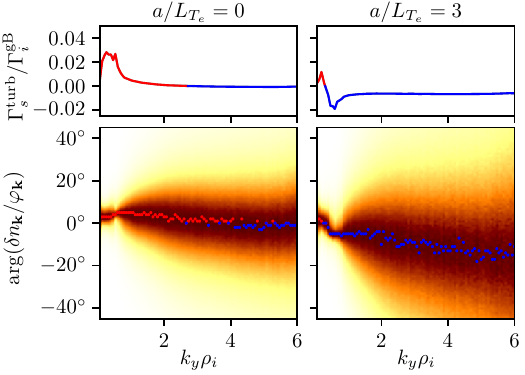}
	\vspace{-5mm}
	\caption{Spectra of the particle flux (top) and phase difference between the density ($\delta n_\mathbf{k}$) and electrostatic potential ($\varphi_\mathbf{k}$) fluctuations (bottom), with $a/L_{T_e}=0$ (left) and $a/L_{T_e}=3$ (right). The dominant phase difference for each $k_y\rho_i$ is denoted by a dot. Positive values are highlighted in red and negative values in blue. }
	\vspace{-2mm}
	\label{fig:phaseshift}
\end{figure} 

Starting with Fig.~\ref{fig:scans}(a), note that turbulence driven solely by either the electron or the ion temperature gradient (being $a/L_{T_i}=3$ or $a/L_{T_e}=3$) cannot lead to peaked density profiles, as convection is outwards $(V>0)$ for the flat density profile cases ($a/L_{n_s}=0$). In contrast, when both temperature gradients coexist, turbulence gives rise to a significant inward convection $(V<0)$, being $\Gamma_s^\text{turb}/\Gamma_i^\text{gB} \approx -0.6$ at a vanishing density gradient. Moreover, the normalized density gradient at which the particle flux is zero is $a/L_{n_s} \approx 0.7$. In other words, if all transport were due to turbulence (and for the fixed values of other parameters used in this plot), the equilibrium density profile would be clearly peaked. 
From Fig.~\ref{fig:scans}(b), it can be observed that a larger electron-temperature-to-ion-temperature ratio enhances the particle pinch. Furthermore, Fig.~\ref{fig:scans}(c) shows that $a/L_{T_i}$ need not be large or comparable to $a/L_{T_e}$ in order to turn the particle flux inwards. The negative convection starts to manifest from values as modest as $a/L_{T_i}\sim 0.25$ when $a/L_{T_e}=3$, and its magnitude increases roughly linearly with the size of $a/L_{T_i}$. 
Similarly,  Fig.~\ref{fig:scans}(d) shows that, although the case with only $a/L_{T_i}=3$ drives positive flux, adding an electron temperature gradient as moderate as $a/L_{T_e}\gtrsim 1.0$, turns the particle flux negative. 

In summary, the existence of an inward turbulent particle flux is very robust and is found over broad regions of the scanned parameter space, whenever the temperature gradients of ions and electrons are both finite and not too small. In Ref.~\cite{Thienpondt_ISHW_2022} it is found that this turbulent pinch holds for different stellarator devices, which aligns with other references that report inward turbulent particle fluxes in LHD \cite{Ishizawa_nf_2017, Nunami_pop_2020}.

Finally, we would like to shed some light on why the electron temperature gradient is key for obtaining inward turbulent particle fluxes. Particle fluxes are driven by density and electrostatic potential fluctuations that are out of phase. For example, if the electrons were treated adiabatically, the density and potential would be in phase, and no particle fluxes would be driven. The sign of the particle flux is thus correlated with the sign of the phase shift between density and potential fluctuations. Figure \ref{fig:phaseshift} shows the spectra (along the normalized binormal wave number $k_y\rho_i$, where $\rho_i$ is the ion Larmor radius)  of the turbulent particle flux (top row) and of the phase difference between density fluctuations, $\delta n_\mathbf{k}$, and electrostatic potential fluctuations, $\varphi_\mathbf{k}$ (bottom row).  The spectra are represented for the case where $a/L_{T_e}=0$ (left column) and $a/L_{T_e}=3$ (right column), both for $a/L_{T_i}=3$ and for $a/L_{n_s}=0$. Inward particle flux contributions occur when the density fluctuations lag behind on the potential fluctuations (negative values of the phase difference), which is what predominantly happens at almost all represented scales for the case with $a/L_{T_e}=3$. In contrast, the phase difference is essentially positive at the scales with largest contributions to the total particle flux when $a/L_{T_e}=0$.


\section{\label{sec:w7xprofiles} Particle balance analysis for a \\ W7-X plasma discharge}

To determine the radial particle flux in an experiment, one needs to compute the neutral ionization source in the confined plasma. In current fusion experiments, the recycling neutrals---i.e., the plasma that is neutralized at the divertor plates and returns to the main plasma---are generally the main source of particles. In Ref.~\cite{Kremeyer_nf_2022} it was concluded that both the main chamber and divertor targets contribute to the fueling of the plasma core in W7-X ECRH plasmas. Once the neutral hydrogen atoms reach the confined plasma they are transported towards the core by charge exchange reactions with the plasma ions, while they ionize by electron impact. 

\begin{figure}[h]
	\includegraphics{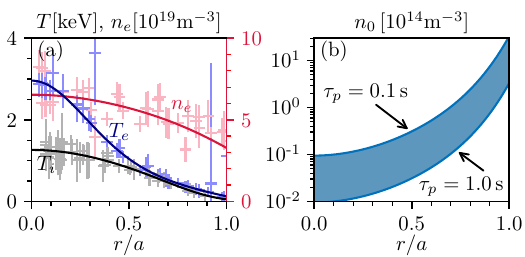} \vspace{-4mm}
	\caption{Plasma profiles (a) and estimated neutral densities (b) for discharge No.~$20180920.017$. }
	\vspace{-5mm}
	\label{fig:profiles}
\end{figure}

The radial profile of the neutral density, $n_0(r)$, and the associated ionization source are estimated using a short-mean-free-path, one-dimensional neutral transport model given by (see, e.g., Ref.~\cite{Hazeltine_nf_1992})
\begin{equation}\label{eq:n0}
\frac{1}{r}\frac{d}{dr}r D_\text{CX}\left(\frac{d n_0}{dr} + \frac{1}{T_0}\frac{dT_0}{dr}n_0\right) = \nu_\text{ion}n_0,
\end{equation}
where $D_\text{CX} = T_0/m_0\nu_\text{CX}$ is the charge exchange diffusivity coefficient, $T_0$ and $m_0$ are the neutral temperature and mass, and $\nu_\text{ion}$ and $\nu_\text{CX}$ are the temperature- and density-dependent ionization and charge exchange frequencies. Neutrals and plasma ions are assumed to be isothermal, $T_0(r) = T_i(r)$, due to frequent charge exchange reactions between them. Figure \ref{fig:profiles}(b) shows the neutral profile obtained with this model for a set of Wendelstein 7-X stationary plasma profiles [Fig.~\ref{fig:profiles}(a)]. 

\vspace{-2mm}
\vspace{0.5cm}

Equation (\ref{eq:n0}) determines the neutral profile up to a multiplicative constant that can be determined from the global particle confinement time $\tau_p$. In Ref.~\cite{Kremeyer_nf_2022}, a confinement time of $\tau_p=0.258$\,s was estimated for a similar discharge using a single-reservoir plasma particle balance. The neutral transport model  has recently been compared with neutral density measurements in W7-X (see Fig.~14 in Ref.~\cite{Beurskens_nf_2021}), showing varying degrees of agreement for $\tau_p$ between 0.15 and 1.0\,s. In Fig.\,\ref{fig:profiles}(b), $\tau_p$ is varied from 0.1 to 1.0\,s to account for uncertainties in this number.

\begin{figure}[h]
	\includegraphics{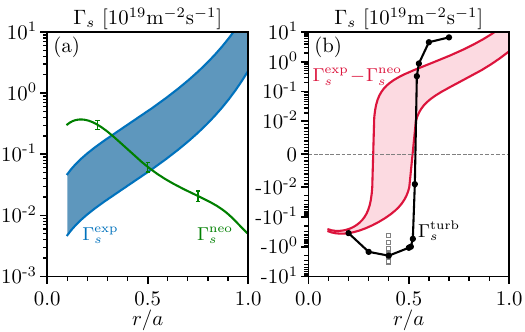} \vspace{-4mm}
	\caption{$\Gamma_s^\text{exp}$, $\Gamma_s^\text{neo}$ and $\Gamma_s^\text{turb}$ for discharge No.~$20180920.017$. The green error bars (a) and the gray squares (b) correspond to $\pm20\%$ variations of the plasma gradients.} \vspace{-2mm}
	\label{fig:gamma}
\end{figure}

Let us denote by $\Gamma_s^{\rm exp}$ the flux profiles calculated, for this range of values of $\tau_p$, by integrating the neutral ionization source $\nu_\text{ion}n_0$ over the plasma radius. The neutral density $n_0$ is obtained from (\ref{eq:n0}) using the experimental profiles shown in Fig.~\ref{fig:profiles}(a).  The results for $\Gamma_s^{\rm exp}$ are displayed in Fig.~\ref{fig:gamma}(a). There is a region in the core (the width of which depends on the assumed $\tau_p$) where the estimated central particle sources are too small to sustain the outward neoclassical radial flux $\Gamma_s^{\rm neo}$ calculated with \texttt{DKES}~\cite{Hirshman1986}. One is led to conclude that an additional particle flux must exist which moves particles towards the magnetic axis in that region. Further outside, on the contrary, the additional flux must be outward directed and account for essentially all the particle flux, since the neoclassical contribution falls one to three orders of magnitude short with respect to $\Gamma_s^{\rm exp}$. A similar shortfall is observed in other stellarators, such as HSX \cite{HSX1, HSX2}.

\vspace{-1mm}
\vspace{1mm}

The turbulent particle flux $\Gamma_s^\text{turb}$ is modeled with \texttt{stella} and shown in Fig.~\ref{fig:gamma}(b) in black. Its sign and size agree with $\Gamma_s^\text{exp} - \Gamma_s^\text{neo}$ reasonably well, taking into account, among other factors, the uncertainties in the plasma profiles and the simplicity of the neutral transport model employed. The effect of the uncertainties of the plasma profiles on the turbulent particle flux is quantified in Appendix B. In addition, in the future, it would be desirable to carry out this kind of analysis using full-flux-surface or radially global gyrokinetic codes.


\vspace{-3mm}
\section{\label{sec:conclusions} Conclusions} 
\vspace{-1mm}

Whereas neoclassical theory predicts strongly hollow density profiles in the plasma core of large stellarators, measured density profiles in Wendelstein 7-X are flat or weakly peaked, indicating that theoretical particle flux calculations are missing an inward contribution. In this Research Letter, we have shown that the discrepancy between theory and experiment disappears when the turbulent component of the particle flux is included in the calculations: Turbulence driven by both the ion and electron temperature gradients provides the missing inward particle flux needed to sustain the experimentally measured density profiles. Moreover, we have performed a one-dimensional particle balance analysis in a Wendelstein 7-X discharge where the sign of the turbulent particle flux agrees, over the entire plasma radius, with the sign of the difference between the experimental and the neoclassical particle fluxes. In particular, the simulations predict a sign change of the turbulent flux, accounting for the fact that the neoclassical flux is too large at the core and too small at the edge of the plasma to explain the experimental particle flux.


\begin{acknowledgments}
	
	\vspace{0.3cm}
	
	The authors acknowledge fruitful discussions with F\'elix I. Parra. This work has been carried out within the framework of the EUROfusion Consortium, funded by the European Union via the Euratom Research and Training Programme (Grant Agreement No.~101052200 -- EUROfusion). Views and opinions expressed are, however, those of the author(s) only and do not necessarily reflect those of the European Union or the European Commission. Neither the European Union nor the European Commission can be held responsible for them. This research was supported in part by Grant No.~PGC2018-095307-B-I00, Ministerio de Ciencia, Innovaci\'on y Universidades, Spain, and by Grant No.~PID2021-123175NB-I00, Ministerio de Ciencia e Innovaci\'on, Spain.  
	
\end{acknowledgments}


\vspace{-2mm}
\section{\label{sec:appendix} Appendix A: Coordinates, resolution and normalization used in stella}
\vspace{-1mm}

The phase-space coordinates employed by \texttt{stella} are denoted by $\{x,y,z,v_\parallel,\mu\}$.  The coordinates perpendicular to the magnetic field lines (Taylor expanded to first order in the flux tube formalism) are defined as $x=r-r_0$ and $y=r_0(\alpha-\alpha_0)$, where $r_0$ and $\alpha_0$ define the center of the flux tube, $r= a \smash{\sqrt{\psi_t/\psi_\text{LCFS}}}$ is the effective minor radius, $a$ is the minor radius, $\psi_t$ is the enclosed toroidal flux divided by $2\pi$ and $\psi_\text{LCFS}$ is the toroidal flux at the last closed flux surface. The field line label is defined as $\alpha = \theta_p - \iota \zeta$, with $\iota$ being the rotational transform and $(\zeta,\theta_p)$ being straight-field line coordinates, which are chosen to be the geometrical toroidal angle $\zeta$ and the poloidal PEST coordinate $\theta_p$ \cite{grimm1983ideal}.  The parallel coordinate $z=\zeta$ is taken to be the geometrical toroidal coordinate. The velocity coordinates are the component of the velocity parallel to the magnetic field, $v_\parallel$, and the magnetic moment, $\mu_s = m_sv^2_\perp/2B$, where $m_s$ is the mass of species $s$, $v_\perp$ is the perpendicular velocity and $B$ is the magnitude of the magnetic field. 

A Fourier decomposition of the electrostatic potential is performed in the directions perpendicular to the magnetic field, for which radial and binormal wave numbers are defined with minimum values $k_{x,\text{min}} = 2\pi/L_x$ and $k_{y,\text{min}} = 2\pi/L_y$, determined by the extent $(L_x, L_y)$ of the perpendicular simulation domain. The simulations are performed with a resolution of $(N_x, N_y, N_z, N_\mu, N_{v_\parallel}) = (91, 271, 49, 12, 48)$. The perpendicular box size in real space is $(L_x, L_y) = (94\rho_i, 94\rho_i)$, which corresponds to a box in Fourier space of $(k_{x,\textrm{max}}\rho_i, k_{y,\textrm{max}}\rho_i) = (2,6)$. Here $\rho_i = v_{{\rm th},i}/\Omega_i$ is the ion Larmor radius with $v_{{\rm th},i} = \smash{\sqrt{2T_i/m_i}}$ being the ion thermal velocity, $\Omega_i = eB_r/m_i$ being the ion gyrofrequency, $T_i$ and $m_i$ being the temperature and mass of the ions, and $B_r = 2\,|\psi_\text{LCFS}|/a^2$ being the reference magnetic field. 
In \texttt{stella}, lengths in the perpendicular direction are normalized with respect to the ion Larmor radius $\rho_i$. The particle flux is normalized with respect to the gyro-Bohm particle flux given by $\Gamma_i^\text{gB} = n_i\,v_{{\rm th},i}\,(\rho_i/a)^2$, where $n_i$ is the density of the ions.  

The generalized twist-and-shift boundary conditions \cite{Martin_ppcf_2018} are employed to deal with the low shear of Wendelstein 7-X. Instead of  relying on the global magnetic shear at a given radial location, we will take advantage of the local magnetic shear, which varies along the field lines. Thanks to this flexibility, the length of the flux tube can be chosen such that the local magnetic shear at the ends of the flux tube imposes $k_{x,\text{min}} = k_{y,\text{min}}$. For the simulations in this Research Letter, the flux tube extends approximately one poloidal turn to ensure an equal spacing in the perpendicular wave number grid.
The flux tube for which the simulations of this Research Letter are run, is commonly known as the bean flux tube, defined by $\alpha_0=0$. It is centered with respect to the equatorial plane, $\theta_p=0$, and the bean-shaped toroidal plane, $\zeta=0$. 

\vspace{-2mm}
\section{\label{sec:appendixb} Appendix B: Uncertainties in the plasma profiles and gradients}
\vspace{-1mm}

\begin{figure}[h]
	\includegraphics[trim={0mm 0mm 0mm 2mm},clip]{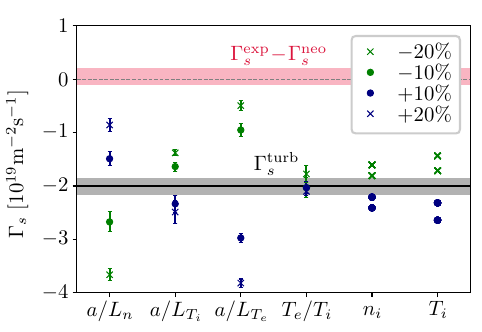} 
	\caption{Effect on the turbulent particle flux of 10 and 20\% changes in the plasma parameters at $r/a=0.4$.  The values of $\Gamma_s^\text{exp}-\Gamma_s^\text{neo}$ and $\Gamma_s^\text{turb}$ shown in Fig.~5 are displayed in red and black, respectively. The error bar on $\Gamma_s^\text{exp}-\Gamma_s^\text{neo}$ represents the uncertainty in $\tau_p$, while the error bar on $\Gamma_s^\text{turb}$ represents the standard mean deviation of the saturated state of the time trace of the simulation.}
	\label{fig:uncertainty}
\end{figure}

To quantify the effect of the uncertainty in the plasma profiles [Fig.\ 4(a)], gyrokinetic simulations are performed with a 10\% (or 20\%) change in the plasma parameters for $r/a=0.4$, the results of which are shown in Fig.~\ref{fig:uncertainty}. Since \texttt{stella} performs simulations with dimensionless parameters, the uncertainties in the plasma density $n_i=n_e$ and the ion temperature $T_i$ are visualized by denormalizing the particle fluxes accordingly. 

\vspace{-1mm}
\vspace{-1mm}

Figure \ref{fig:uncertainty} suggests that at $r/a=0.4$, higher density gradients and lower temperature gradients would improve the agreement between $\Gamma_s^\text{exp}-\Gamma_s^\text{neo}$ and $\Gamma_s^\text{turb}$. Similarly, lower densities and temperatures would improve agreement, although their influence is smaller compared with the effect of changing the gradients.


\bibliography{bibliography.bib}

\begin{thebibliography}{24}%
\makeatletter
\providecommand \@ifxundefined [1]{%
 \@ifx{#1\undefined}
}%
\providecommand \@ifnum [1]{%
 \ifnum #1\expandafter \@firstoftwo
 \else \expandafter \@secondoftwo
 \fi
}%
\providecommand \@ifx [1]{%
 \ifx #1\expandafter \@firstoftwo
 \else \expandafter \@secondoftwo
 \fi
}%
\providecommand \natexlab [1]{#1}%
\providecommand \enquote  [1]{``#1''}%
\providecommand \bibnamefont  [1]{#1}%
\providecommand \bibfnamefont [1]{#1}%
\providecommand \citenamefont [1]{#1}%
\providecommand \href@noop [0]{\@secondoftwo}%
\providecommand \href [0]{\begingroup \@sanitize@url \@href}%
\providecommand \@href[1]{\@@startlink{#1}\@@href}%
\providecommand \@@href[1]{\endgroup#1\@@endlink}%
\providecommand \@sanitize@url [0]{\catcode `\\12\catcode `\$12\catcode
  `\&12\catcode `\#12\catcode `\^12\catcode `\_12\catcode `\%12\relax}%
\providecommand \@@startlink[1]{}%
\providecommand \@@endlink[0]{}%
\providecommand \url  [0]{\begingroup\@sanitize@url \@url }%
\providecommand \@url [1]{\endgroup\@href {#1}{\urlprefix }}%
\providecommand \urlprefix  [0]{URL }%
\providecommand \Eprint [0]{\href }%
\providecommand \doibase [0]{http://dx.doi.org/}%
\providecommand \selectlanguage [0]{\@gobble}%
\providecommand \bibinfo  [0]{\@secondoftwo}%
\providecommand \bibfield  [0]{\@secondoftwo}%
\providecommand \translation [1]{[#1]}%
\providecommand \BibitemOpen [0]{}%
\providecommand \bibitemStop [0]{}%
\providecommand \bibitemNoStop [0]{.\EOS\space}%
\providecommand \EOS [0]{\spacefactor3000\relax}%
\providecommand \BibitemShut  [1]{\csname bibitem#1\endcsname}%
\let\auto@bib@innerbib\@empty
\bibitem [{\citenamefont {Grieger}\ \emph {et~al.}(1992)\citenamefont
  {Grieger}, \citenamefont {Beidler}, \citenamefont {Harmeyer}, \citenamefont
  {Lotz}, \citenamefont {Ki{\ss}Linger}, \citenamefont {Merkel}, \citenamefont
  {N\"{u}hrenberg}, \citenamefont {Rau}, \citenamefont {Strumberger},\ and\
  \citenamefont {Wobig}}]{Grieger_fst_1992}%
  \BibitemOpen
  \bibfield  {author} {\bibinfo {author} {\bibfnamefont {G.}~\bibnamefont
  {Grieger}}, \bibinfo {author} {\bibfnamefont {C.}~\bibnamefont {Beidler}},
  \bibinfo {author} {\bibfnamefont {E.}~\bibnamefont {Harmeyer}}, \bibinfo
  {author} {\bibfnamefont {W.}~\bibnamefont {Lotz}}, \bibinfo {author}
  {\bibfnamefont {J.}~\bibnamefont {Ki{\ss}Linger}}, \bibinfo {author}
  {\bibfnamefont {P.}~\bibnamefont {Merkel}}, \bibinfo {author} {\bibfnamefont
  {J.}~\bibnamefont {N\"{u}hrenberg}}, \bibinfo {author} {\bibfnamefont
  {F.}~\bibnamefont {Rau}}, \bibinfo {author} {\bibfnamefont {E.}~\bibnamefont
  {Strumberger}}, \ and\ \bibinfo {author} {\bibfnamefont {H.}~\bibnamefont
  {Wobig}},\ }\href {\doibase 10.13182/fst92-a29977} {\bibfield  {journal}
  {\bibinfo  {journal} {Fusion Technology}\ }\textbf {\bibinfo {volume} {21}},\
  \bibinfo {pages} {1767} (\bibinfo {year} {1992})}\BibitemShut {NoStop}%
\bibitem [{\citenamefont {Pedersen}\ \emph {et~al.}(2016)\citenamefont
  {Pedersen}, \citenamefont {Otte}, \citenamefont {Lazerson}, \citenamefont
  {Helander}, \citenamefont {Bozhenkov}, \citenamefont {Biedermann},
  \citenamefont {Klinger}, \citenamefont {Wolf},\ and\ \citenamefont
  {Bosch}}]{Pedersen_nature_2016}%
  \BibitemOpen
  \bibfield  {author} {\bibinfo {author} {\bibfnamefont {T.~S.}\ \bibnamefont
  {Pedersen}}, \bibinfo {author} {\bibfnamefont {M.}~\bibnamefont {Otte}},
  \bibinfo {author} {\bibfnamefont {S.}~\bibnamefont {Lazerson}}, \bibinfo
  {author} {\bibfnamefont {P.}~\bibnamefont {Helander}}, \bibinfo {author}
  {\bibfnamefont {S.}~\bibnamefont {Bozhenkov}}, \bibinfo {author}
  {\bibfnamefont {C.}~\bibnamefont {Biedermann}}, \bibinfo {author}
  {\bibfnamefont {T.}~\bibnamefont {Klinger}}, \bibinfo {author} {\bibfnamefont
  {R.~C.}\ \bibnamefont {Wolf}}, \ and\ \bibinfo {author} {\bibfnamefont
  {H.~S.}\ \bibnamefont {Bosch}},\ }\href {https://doi.org/10.1038/ncomms13493}
  {\bibfield  {journal} {\bibinfo  {journal} {Nature Communications}\ }\textbf
  {\bibinfo {volume} {7}} (\bibinfo {year} {2016})}\BibitemShut {NoStop}%
\bibitem [{\citenamefont {Beidler}\ \emph {et~al.}(2021)\citenamefont
  {Beidler}, \citenamefont {Smith}, \citenamefont {Alonso}, \citenamefont
  {Andreeva}, \citenamefont {Baldzuhn}, \citenamefont {Beurskens},
  \citenamefont {Borchardt}, \citenamefont {Bozhenkov}, \citenamefont
  {Brunner}, \citenamefont {Damm}, \citenamefont {Drevlak}, \citenamefont
  {Ford}, \citenamefont {Fuchert}, \citenamefont {Geiger}, \citenamefont
  {Helander}, \citenamefont {Hergenhahn}, \citenamefont {Hirsch}, \citenamefont
  {H\"{o}fel}, \citenamefont {Kazakov}, \citenamefont {Kleiber}, \citenamefont
  {Krychowiak}, \citenamefont {Kwak}, \citenamefont {Langenberg}, \citenamefont
  {Laqua}, \citenamefont {Neuner}, \citenamefont {Pablant}, \citenamefont
  {Pasch}, \citenamefont {Pavone}, \citenamefont {Pedersen}, \citenamefont
  {Rahbarnia}, \citenamefont {Schilling}, \citenamefont {Scott}, \citenamefont
  {Stange}, \citenamefont {Svensson}, \citenamefont {Thomsen}, \citenamefont
  {Turkin}, \citenamefont {Warmer}, \citenamefont {Wolf}, \citenamefont
  {Zhang}, \citenamefont {Abramovic}, \citenamefont {\"{A}k\"{a}slompolo},
  \citenamefont {Alcus{\'{o}}n}, \citenamefont {Aleynikov}, \citenamefont
  {Aleynikova}, \citenamefont {Ali}, \citenamefont {Alonso}, \citenamefont
  {Anda}, \citenamefont {Ascasibar}, \citenamefont {B\"{a}hner}, \citenamefont
  {Baek}, \citenamefont {Balden}, \citenamefont {Banduch}, \citenamefont
  {Barbui}, \citenamefont {Behr}, \citenamefont {Benndorf}, \citenamefont
  {Biedermann}, \citenamefont {Biel}, \citenamefont {Blackwell}, \citenamefont
  {Blanco}, \citenamefont {Blatzheim}, \citenamefont {Ballinger}, \citenamefont
  {Bluhm}, \citenamefont {B\"{o}ckenhoff}, \citenamefont {B\"{o}swirth},
  \citenamefont {B\"{o}ttger}, \citenamefont {Borsuk}, \citenamefont {Boscary},
  \citenamefont {Bosch}, \citenamefont {Brakel}, \citenamefont {Brand},
  \citenamefont {Brandt}, \citenamefont {Br\"{a}uer}, \citenamefont {Braune},
  \citenamefont {Brezinsek}, \citenamefont {Brunner}, \citenamefont {Burhenn},
  \citenamefont {Bussiahn}, \citenamefont {Buttensch\"{o}n}, \citenamefont
  {Bykov}, \citenamefont {Cai}, \citenamefont {Calvo}, \citenamefont {Cannas},
  \citenamefont {Cappa}, \citenamefont {Carls}, \citenamefont {Carraro},
  \citenamefont {Carvalho}, \citenamefont {Castejon}, \citenamefont {Charl},
  \citenamefont {Chaudhary}, \citenamefont {Chauvin}, \citenamefont
  {Chernyshev}, \citenamefont {Cianciosa}, \citenamefont {Citarella},
  \citenamefont {Claps}, \citenamefont {Coenen}, \citenamefont {Cole},
  \citenamefont {Cole}, \citenamefont {Cordella}, \citenamefont {Cseh},
  \citenamefont {Czarnecka}, \citenamefont {Czerski}, \citenamefont
  {Czerwinski}, \citenamefont {Czymek}, \citenamefont {da~Molin}, \citenamefont
  {da~Silva}, \citenamefont {de~la Pena}, \citenamefont {Degenkolbe},
  \citenamefont {Dhard}, \citenamefont {Dibon}, \citenamefont {Dinklage},
  \citenamefont {Dittmar}, \citenamefont {Drewelow}, \citenamefont {Drews},
  \citenamefont {Durodie}, \citenamefont {Edlund}, \citenamefont {Effenberg},
  \citenamefont {Ehrke}, \citenamefont {Elgeti}, \citenamefont {Endler},
  \citenamefont {Ennis}, \citenamefont {Esteban}, \citenamefont {Estrada},
  \citenamefont {Fellinger}, \citenamefont {Feng}, \citenamefont {Flom},
  \citenamefont {Fernandes}, \citenamefont {Fietz}, \citenamefont {Figacz},
  \citenamefont {Fontdecaba}, \citenamefont {Fornal}, \citenamefont {Frerichs},
  \citenamefont {Freund}, \citenamefont {Funaba}, \citenamefont {Galkowski},
  \citenamefont {Gantenbein}, \citenamefont {Gao}, \citenamefont
  {Rega{\~{n}}a}, \citenamefont {Gates}, \citenamefont {Geiger}, \citenamefont
  {Giannella}, \citenamefont {Gogoleva}, \citenamefont {Goncalves},
  \citenamefont {Goriaev}, \citenamefont {Gradic}, \citenamefont {Grahl},
  \citenamefont {Green}, \citenamefont {Greuner}, \citenamefont {Grosman},
  \citenamefont {Grote}, \citenamefont {Gruca}, \citenamefont {Grulke},
  \citenamefont {Guerard}, \citenamefont {Hacker}, \citenamefont {Han},
  \citenamefont {Harris}, \citenamefont {Hartmann}, \citenamefont
  {Hathiramani}, \citenamefont {Hein}, \citenamefont {Heinemann}, \citenamefont
  {Helander}, \citenamefont {Henneberg}, \citenamefont {Henkel}, \citenamefont
  {Hergenhahn}, \citenamefont {Sanchez}, \citenamefont {Hidalgo}, \citenamefont
  {Hollfeld}, \citenamefont {H\"{o}lting}, \citenamefont {H\"{o}schen},
  \citenamefont {Houry}, \citenamefont {Howard}, \citenamefont {Huang},
  \citenamefont {Huang}, \citenamefont {Hubeny}, \citenamefont {Huber},
  \citenamefont {Hunger}, \citenamefont {Ida}, \citenamefont {Ilkei},
  \citenamefont {Illy}, \citenamefont {Israeli}, \citenamefont {Jablonski},
  \citenamefont {Jakubowski}, \citenamefont {Jelonnek}, \citenamefont
  {Jenzsch}, \citenamefont {Jesche}, \citenamefont {Jia}, \citenamefont
  {Junghanns}, \citenamefont {Kacmarczyk}, \citenamefont {Kallmeyer},
  \citenamefont {Kamionka}, \citenamefont {Kasahara}, \citenamefont {Kasparek},
  \citenamefont {Kenmochi}, \citenamefont {Killer}, \citenamefont {Kirschner},
  \citenamefont {Klinger}, \citenamefont {Knauer}, \citenamefont {Knaup},
  \citenamefont {Knieps}, \citenamefont {Kobarg}, \citenamefont {Kocsis},
  \citenamefont {K\"{o}chl}, \citenamefont {Kolesnichenko}, \citenamefont
  {K\"{o}nies}, \citenamefont {K\"{o}nig}, \citenamefont {Kornejew},
  \citenamefont {Koschinsky}, \citenamefont {K\"{o}ster}, \citenamefont
  {Kr\"{a}mer}, \citenamefont {Krampitz}, \citenamefont {Kr\"{a}mer-Flecken},
  \citenamefont {Krawczyk}, \citenamefont {Kremeyer}, \citenamefont {Krom},
  \citenamefont {Ksiazek}, \citenamefont {Kubkowska}, \citenamefont
  {K\"{u}hner}, \citenamefont {Kurki-Suonio}, \citenamefont {Kurz},
  \citenamefont {Landreman}, \citenamefont {Lang}, \citenamefont {Lang},
  \citenamefont {Langish}, \citenamefont {Laqua}, \citenamefont {Laube},
  \citenamefont {Lazerson}, \citenamefont {Lechte}, \citenamefont {Lennartz},
  \citenamefont {Leonhardt}, \citenamefont {Li}, \citenamefont {Li},
  \citenamefont {Li}, \citenamefont {Liang}, \citenamefont {Linsmeier},
  \citenamefont {Liu}, \citenamefont {Lobsien}, \citenamefont {Loesser},
  \citenamefont {Cisquella}, \citenamefont {Lore}, \citenamefont {Lorenz},
  \citenamefont {Losert}, \citenamefont {L\"{u}cke}, \citenamefont {Lumsdaine},
  \citenamefont {Lutsenko}, \citenamefont {Maa{\ss}berg}, \citenamefont
  {Marchuk}, \citenamefont {Matthew}, \citenamefont {Marsen}, \citenamefont
  {Marushchenko}, \citenamefont {Masuzaki}, \citenamefont {Maurer},
  \citenamefont {Mayer}, \citenamefont {McCarthy}, \citenamefont {McNeely},
  \citenamefont {Meier}, \citenamefont {Mellein}, \citenamefont {Mendelevitch},
  \citenamefont {Mertens}, \citenamefont {Mikkelsen}, \citenamefont
  {Mishchenko}, \citenamefont {Missal}, \citenamefont {Mittelstaedt},
  \citenamefont {Mizuuchi}, \citenamefont {Mollen}, \citenamefont {Moncada},
  \citenamefont {M\"{o}nnich}, \citenamefont {Morisaki}, \citenamefont
  {Moseev}, \citenamefont {Murakami}, \citenamefont {N{\'{a}}fr{\'{a}}di},
  \citenamefont {Nagel}, \citenamefont {Naujoks}, \citenamefont {Neilson},
  \citenamefont {Neu}, \citenamefont {Neubauer}, \citenamefont {Ngo},
  \citenamefont {Nicolai}, \citenamefont {Nielsen}, \citenamefont {Niemann},
  \citenamefont {Nishizawa}, \citenamefont {Nocentini}, \citenamefont
  {N\"{u}hrenberg}, \citenamefont {N\"{u}hrenberg}, \citenamefont {Obermayer},
  \citenamefont {Offermanns}, \citenamefont {Ogawa}, \citenamefont
  {\"{O}lmanns}, \citenamefont {Ongena}, \citenamefont {Oosterbeek},
  \citenamefont {Orozco}, \citenamefont {Otte}, \citenamefont {Rodriguez},
  \citenamefont {Panadero}, \citenamefont {Alvarez}, \citenamefont
  {Papenfu{\ss}}, \citenamefont {Paqay}, \citenamefont {Pawelec}, \citenamefont
  {Pedersen}, \citenamefont {Pelka}, \citenamefont {Perseo}, \citenamefont
  {Peterson}, \citenamefont {Pilopp}, \citenamefont {Pingel}, \citenamefont
  {Pisano}, \citenamefont {Plaum}, \citenamefont {Plunk}, \citenamefont
  {P\"{o}l\"{o}skei}, \citenamefont {Porkolab}, \citenamefont {Proll},
  \citenamefont {Puiatti}, \citenamefont {Sitjes}, \citenamefont {Purps},
  \citenamefont {Rack}, \citenamefont {R{\'{e}}csei}, \citenamefont {Reiman},
  \citenamefont {Reimold}, \citenamefont {Reiter}, \citenamefont {Remppel},
  \citenamefont {Renard}, \citenamefont {Riedl}, \citenamefont {Riemann},
  \citenamefont {Risse}, \citenamefont {Rohde}, \citenamefont {R\"{o}hlinger},
  \citenamefont {Rom{\'{e}}}, \citenamefont {Rondeshagen}, \citenamefont
  {Rong}, \citenamefont {Roth}, \citenamefont {Rudischhauser}, \citenamefont
  {Rummel}, \citenamefont {Rummel}, \citenamefont {Runov}, \citenamefont
  {Rust}, \citenamefont {Ryc}, \citenamefont {Ryosuke}, \citenamefont
  {Sakamoto}, \citenamefont {Salewski}, \citenamefont {Samartsev},
  \citenamefont {S{\'{a}}nchez}, \citenamefont {Sano}, \citenamefont {Satake},
  \citenamefont {Schacht}, \citenamefont {Satheeswaran}, \citenamefont
  {Schauer}, \citenamefont {Scherer}, \citenamefont {Schlaich}, \citenamefont
  {Schlisio}, \citenamefont {Schluck}, \citenamefont {Schl\"{u}ter},
  \citenamefont {Schmitt}, \citenamefont {Schmitz}, \citenamefont {Schmitz},
  \citenamefont {Schmuck}, \citenamefont {Schneider}, \citenamefont
  {Schneider}, \citenamefont {Scholz}, \citenamefont {Schrittwieser},
  \citenamefont {Schr\"{o}der}, \citenamefont {Schr\"{o}der}, \citenamefont
  {Schroeder}, \citenamefont {Schumacher}, \citenamefont {Schweer},
  \citenamefont {Sereda}, \citenamefont {Shanahan}, \citenamefont {Sibilia},
  \citenamefont {Sinha}, \citenamefont {Sipli\"{a}}, \citenamefont {Slaby},
  \citenamefont {Sleczka}, \citenamefont {Spiess}, \citenamefont {Spong},
  \citenamefont {Spring}, \citenamefont {Stadler}, \citenamefont {Stejner},
  \citenamefont {Stephey}, \citenamefont {Stridde}, \citenamefont {Suzuki},
  \citenamefont {Szab{\'{o}}}, \citenamefont {Szabolics}, \citenamefont
  {Szepesi}, \citenamefont {Sz\"{o}kefalvi-Nagy}, \citenamefont {Tamura},
  \citenamefont {Tancetti}, \citenamefont {Terry}, \citenamefont {Thomas},
  \citenamefont {Thumm}, \citenamefont {Travere}, \citenamefont {Traverso},
  \citenamefont {Tretter}, \citenamefont {Mora}, \citenamefont {Tsuchiya},
  \citenamefont {Tsujimura}, \citenamefont {Tulip{\'{a}}n}, \citenamefont
  {Unterberg}, \citenamefont {Vakulchyk}, \citenamefont {Valet}, \citenamefont
  {Van{\'{o}}}, \citenamefont {van Eeten}, \citenamefont {van Milligen},
  \citenamefont {van Vuuren}, \citenamefont {Vela}, \citenamefont {Velasco},
  \citenamefont {Vergote}, \citenamefont {Vervier}, \citenamefont {Vianello},
  \citenamefont {Viebke}, \citenamefont {Vilbrandt}, \citenamefont {von
  Stechow}, \citenamefont {Vork\"{o}per}, \citenamefont {Wadle}, \citenamefont
  {Wagner}, \citenamefont {Wang}, \citenamefont {Wang}, \citenamefont {Wang},
  \citenamefont {Wauters}, \citenamefont {Wegener}, \citenamefont {Weggen},
  \citenamefont {Wegner}, \citenamefont {Wei}, \citenamefont {Weir},
  \citenamefont {Wendorf}, \citenamefont {Wenzel}, \citenamefont {Werner},
  \citenamefont {White}, \citenamefont {Wiegel}, \citenamefont {Wilde},
  \citenamefont {Windisch}, \citenamefont {Winkler}, \citenamefont {Winter},
  \citenamefont {Winters}, \citenamefont {Wolf}, \citenamefont {Wolf},
  \citenamefont {Wright}, \citenamefont {Wurden}, \citenamefont {Xanthopoulos},
  \citenamefont {Yamada}, \citenamefont {Yamada}, \citenamefont {Yasuhara},
  \citenamefont {Yokoyama}, \citenamefont {Zanini}, \citenamefont {Zarnstorff},
  \citenamefont {Zeitler}, \citenamefont {Zhang}, \citenamefont {Zhu},
  \citenamefont {Zilker}, \citenamefont {Zocco}, \citenamefont {Zoletnik},\
  and\ \citenamefont {and}}]{Beidler_nature_2021}%
  \BibitemOpen
  \bibfield  {author} {\bibinfo {author} {\bibfnamefont {C.~D.}\ \bibnamefont
  {Beidler}}, \bibinfo {author} {\bibfnamefont {H.~M.}\ \bibnamefont {Smith}},
  \bibinfo {author} {\bibfnamefont {A.}~\bibnamefont {Alonso}}, \bibinfo
  {author} {\bibfnamefont {T.}~\bibnamefont {Andreeva}}, \bibinfo {author}
  {\bibfnamefont {J.}~\bibnamefont {Baldzuhn}}, \bibinfo {author}
  {\bibfnamefont {M.~N.~A.}\ \bibnamefont {Beurskens}}, \bibinfo {author}
  {\bibfnamefont {M.}~\bibnamefont {Borchardt}}, \bibinfo {author}
  {\bibfnamefont {S.~A.}\ \bibnamefont {Bozhenkov}}, \bibinfo {author}
  {\bibfnamefont {K.~J.}\ \bibnamefont {Brunner}}, \bibinfo {author}
  {\bibfnamefont {H.}~\bibnamefont {Damm}}, \bibinfo {author} {\bibfnamefont
  {M.}~\bibnamefont {Drevlak}}, \bibinfo {author} {\bibfnamefont {O.~P.}\
  \bibnamefont {Ford}}, \bibinfo {author} {\bibfnamefont {G.}~\bibnamefont
  {Fuchert}}, \bibinfo {author} {\bibfnamefont {J.}~\bibnamefont {Geiger}},
  \bibinfo {author} {\bibfnamefont {P.}~\bibnamefont {Helander}}, \bibinfo
  {author} {\bibfnamefont {U.}~\bibnamefont {Hergenhahn}}, \bibinfo {author}
  {\bibfnamefont {M.}~\bibnamefont {Hirsch}}, \bibinfo {author} {\bibfnamefont
  {U.}~\bibnamefont {H\"{o}fel}}, \bibinfo {author} {\bibfnamefont {Y.~O.}\
  \bibnamefont {Kazakov}}, \bibinfo {author} {\bibfnamefont {R.}~\bibnamefont
  {Kleiber}}, \bibinfo {author} {\bibfnamefont {M.}~\bibnamefont {Krychowiak}},
  \bibinfo {author} {\bibfnamefont {S.}~\bibnamefont {Kwak}}, \bibinfo {author}
  {\bibfnamefont {A.}~\bibnamefont {Langenberg}}, \bibinfo {author}
  {\bibfnamefont {H.~P.}\ \bibnamefont {Laqua}}, \bibinfo {author}
  {\bibfnamefont {U.}~\bibnamefont {Neuner}}, \bibinfo {author} {\bibfnamefont
  {N.~A.}\ \bibnamefont {Pablant}}, \bibinfo {author} {\bibfnamefont
  {E.}~\bibnamefont {Pasch}}, \bibinfo {author} {\bibfnamefont
  {A.}~\bibnamefont {Pavone}}, \bibinfo {author} {\bibfnamefont {T.~S.}\
  \bibnamefont {Pedersen}}, \bibinfo {author} {\bibfnamefont {K.}~\bibnamefont
  {Rahbarnia}}, \bibinfo {author} {\bibfnamefont {J.}~\bibnamefont
  {Schilling}}, \bibinfo {author} {\bibfnamefont {E.~R.}\ \bibnamefont
  {Scott}}, \bibinfo {author} {\bibfnamefont {T.}~\bibnamefont {Stange}},
  \bibinfo {author} {\bibfnamefont {J.}~\bibnamefont {Svensson}}, \bibinfo
  {author} {\bibfnamefont {H.}~\bibnamefont {Thomsen}}, \bibinfo {author}
  {\bibfnamefont {Y.}~\bibnamefont {Turkin}}, \bibinfo {author} {\bibfnamefont
  {F.}~\bibnamefont {Warmer}}, \bibinfo {author} {\bibfnamefont {R.~C.}\
  \bibnamefont {Wolf}}, \bibinfo {author} {\bibfnamefont {D.}~\bibnamefont
  {Zhang}}, \bibinfo {author} {\bibfnamefont {I.}~\bibnamefont {Abramovic}},
  \bibinfo {author} {\bibfnamefont {S.}~\bibnamefont {\"{A}k\"{a}slompolo}},
  \bibinfo {author} {\bibfnamefont {J.}~\bibnamefont {Alcus{\'{o}}n}}, \bibinfo
  {author} {\bibfnamefont {P.}~\bibnamefont {Aleynikov}}, \bibinfo {author}
  {\bibfnamefont {K.}~\bibnamefont {Aleynikova}}, \bibinfo {author}
  {\bibfnamefont {A.}~\bibnamefont {Ali}}, \bibinfo {author} {\bibfnamefont
  {A.}~\bibnamefont {Alonso}}, \bibinfo {author} {\bibfnamefont
  {G.}~\bibnamefont {Anda}}, \bibinfo {author} {\bibfnamefont {E.}~\bibnamefont
  {Ascasibar}}, \bibinfo {author} {\bibfnamefont {J.~P.}\ \bibnamefont
  {B\"{a}hner}}, \bibinfo {author} {\bibfnamefont {S.~G.}\ \bibnamefont
  {Baek}}, \bibinfo {author} {\bibfnamefont {M.}~\bibnamefont {Balden}},
  \bibinfo {author} {\bibfnamefont {M.}~\bibnamefont {Banduch}}, \bibinfo
  {author} {\bibfnamefont {T.}~\bibnamefont {Barbui}}, \bibinfo {author}
  {\bibfnamefont {W.}~\bibnamefont {Behr}}, \bibinfo {author} {\bibfnamefont
  {A.}~\bibnamefont {Benndorf}}, \bibinfo {author} {\bibfnamefont
  {C.}~\bibnamefont {Biedermann}}, \bibinfo {author} {\bibfnamefont
  {W.}~\bibnamefont {Biel}}, \bibinfo {author} {\bibfnamefont {B.}~\bibnamefont
  {Blackwell}}, \bibinfo {author} {\bibfnamefont {E.}~\bibnamefont {Blanco}},
  \bibinfo {author} {\bibfnamefont {M.}~\bibnamefont {Blatzheim}}, \bibinfo
  {author} {\bibfnamefont {S.}~\bibnamefont {Ballinger}}, \bibinfo {author}
  {\bibfnamefont {T.}~\bibnamefont {Bluhm}}, \bibinfo {author} {\bibfnamefont
  {D.}~\bibnamefont {B\"{o}ckenhoff}}, \bibinfo {author} {\bibfnamefont
  {B.}~\bibnamefont {B\"{o}swirth}}, \bibinfo {author} {\bibfnamefont {L.-G.}\
  \bibnamefont {B\"{o}ttger}}, \bibinfo {author} {\bibfnamefont
  {V.}~\bibnamefont {Borsuk}}, \bibinfo {author} {\bibfnamefont
  {J.}~\bibnamefont {Boscary}}, \bibinfo {author} {\bibfnamefont {H.-S.}\
  \bibnamefont {Bosch}}, \bibinfo {author} {\bibfnamefont {R.}~\bibnamefont
  {Brakel}}, \bibinfo {author} {\bibfnamefont {H.}~\bibnamefont {Brand}},
  \bibinfo {author} {\bibfnamefont {C.}~\bibnamefont {Brandt}}, \bibinfo
  {author} {\bibfnamefont {T.}~\bibnamefont {Br\"{a}uer}}, \bibinfo {author}
  {\bibfnamefont {H.}~\bibnamefont {Braune}}, \bibinfo {author} {\bibfnamefont
  {S.}~\bibnamefont {Brezinsek}}, \bibinfo {author} {\bibfnamefont {K.-J.}\
  \bibnamefont {Brunner}}, \bibinfo {author} {\bibfnamefont {R.}~\bibnamefont
  {Burhenn}}, \bibinfo {author} {\bibfnamefont {R.}~\bibnamefont {Bussiahn}},
  \bibinfo {author} {\bibfnamefont {B.}~\bibnamefont {Buttensch\"{o}n}},
  \bibinfo {author} {\bibfnamefont {V.}~\bibnamefont {Bykov}}, \bibinfo
  {author} {\bibfnamefont {J.}~\bibnamefont {Cai}}, \bibinfo {author}
  {\bibfnamefont {I.}~\bibnamefont {Calvo}}, \bibinfo {author} {\bibfnamefont
  {B.}~\bibnamefont {Cannas}}, \bibinfo {author} {\bibfnamefont
  {A.}~\bibnamefont {Cappa}}, \bibinfo {author} {\bibfnamefont
  {A.}~\bibnamefont {Carls}}, \bibinfo {author} {\bibfnamefont
  {L.}~\bibnamefont {Carraro}}, \bibinfo {author} {\bibfnamefont
  {B.}~\bibnamefont {Carvalho}}, \bibinfo {author} {\bibfnamefont
  {F.}~\bibnamefont {Castejon}}, \bibinfo {author} {\bibfnamefont
  {A.}~\bibnamefont {Charl}}, \bibinfo {author} {\bibfnamefont
  {N.}~\bibnamefont {Chaudhary}}, \bibinfo {author} {\bibfnamefont
  {D.}~\bibnamefont {Chauvin}}, \bibinfo {author} {\bibfnamefont
  {F.}~\bibnamefont {Chernyshev}}, \bibinfo {author} {\bibfnamefont
  {M.}~\bibnamefont {Cianciosa}}, \bibinfo {author} {\bibfnamefont
  {R.}~\bibnamefont {Citarella}}, \bibinfo {author} {\bibfnamefont
  {G.}~\bibnamefont {Claps}}, \bibinfo {author} {\bibfnamefont
  {J.}~\bibnamefont {Coenen}}, \bibinfo {author} {\bibfnamefont
  {M.}~\bibnamefont {Cole}}, \bibinfo {author} {\bibfnamefont {M.~J.}\
  \bibnamefont {Cole}}, \bibinfo {author} {\bibfnamefont {F.}~\bibnamefont
  {Cordella}}, \bibinfo {author} {\bibfnamefont {G.}~\bibnamefont {Cseh}},
  \bibinfo {author} {\bibfnamefont {A.}~\bibnamefont {Czarnecka}}, \bibinfo
  {author} {\bibfnamefont {K.}~\bibnamefont {Czerski}}, \bibinfo {author}
  {\bibfnamefont {M.}~\bibnamefont {Czerwinski}}, \bibinfo {author}
  {\bibfnamefont {G.}~\bibnamefont {Czymek}}, \bibinfo {author} {\bibfnamefont
  {A.}~\bibnamefont {da~Molin}}, \bibinfo {author} {\bibfnamefont
  {A.}~\bibnamefont {da~Silva}}, \bibinfo {author} {\bibfnamefont
  {A.}~\bibnamefont {de~la Pena}}, \bibinfo {author} {\bibfnamefont
  {S.}~\bibnamefont {Degenkolbe}}, \bibinfo {author} {\bibfnamefont {C.~P.}\
  \bibnamefont {Dhard}}, \bibinfo {author} {\bibfnamefont {M.}~\bibnamefont
  {Dibon}}, \bibinfo {author} {\bibfnamefont {A.}~\bibnamefont {Dinklage}},
  \bibinfo {author} {\bibfnamefont {T.}~\bibnamefont {Dittmar}}, \bibinfo
  {author} {\bibfnamefont {P.}~\bibnamefont {Drewelow}}, \bibinfo {author}
  {\bibfnamefont {P.}~\bibnamefont {Drews}}, \bibinfo {author} {\bibfnamefont
  {F.}~\bibnamefont {Durodie}}, \bibinfo {author} {\bibfnamefont
  {E.}~\bibnamefont {Edlund}}, \bibinfo {author} {\bibfnamefont
  {F.}~\bibnamefont {Effenberg}}, \bibinfo {author} {\bibfnamefont
  {G.}~\bibnamefont {Ehrke}}, \bibinfo {author} {\bibfnamefont
  {S.}~\bibnamefont {Elgeti}}, \bibinfo {author} {\bibfnamefont
  {M.}~\bibnamefont {Endler}}, \bibinfo {author} {\bibfnamefont
  {D.}~\bibnamefont {Ennis}}, \bibinfo {author} {\bibfnamefont
  {H.}~\bibnamefont {Esteban}}, \bibinfo {author} {\bibfnamefont
  {T.}~\bibnamefont {Estrada}}, \bibinfo {author} {\bibfnamefont
  {J.}~\bibnamefont {Fellinger}}, \bibinfo {author} {\bibfnamefont
  {Y.}~\bibnamefont {Feng}}, \bibinfo {author} {\bibfnamefont {E.}~\bibnamefont
  {Flom}}, \bibinfo {author} {\bibfnamefont {H.}~\bibnamefont {Fernandes}},
  \bibinfo {author} {\bibfnamefont {W.~H.}\ \bibnamefont {Fietz}}, \bibinfo
  {author} {\bibfnamefont {W.}~\bibnamefont {Figacz}}, \bibinfo {author}
  {\bibfnamefont {J.}~\bibnamefont {Fontdecaba}}, \bibinfo {author}
  {\bibfnamefont {T.}~\bibnamefont {Fornal}}, \bibinfo {author} {\bibfnamefont
  {H.}~\bibnamefont {Frerichs}}, \bibinfo {author} {\bibfnamefont
  {A.}~\bibnamefont {Freund}}, \bibinfo {author} {\bibfnamefont
  {T.}~\bibnamefont {Funaba}}, \bibinfo {author} {\bibfnamefont
  {A.}~\bibnamefont {Galkowski}}, \bibinfo {author} {\bibfnamefont
  {G.}~\bibnamefont {Gantenbein}}, \bibinfo {author} {\bibfnamefont
  {Y.}~\bibnamefont {Gao}}, \bibinfo {author} {\bibfnamefont {J.~G.}\
  \bibnamefont {Rega{\~{n}}a}}, \bibinfo {author} {\bibfnamefont
  {D.}~\bibnamefont {Gates}}, \bibinfo {author} {\bibfnamefont
  {B.}~\bibnamefont {Geiger}}, \bibinfo {author} {\bibfnamefont
  {V.}~\bibnamefont {Giannella}}, \bibinfo {author} {\bibfnamefont
  {A.}~\bibnamefont {Gogoleva}}, \bibinfo {author} {\bibfnamefont
  {B.}~\bibnamefont {Goncalves}}, \bibinfo {author} {\bibfnamefont
  {A.}~\bibnamefont {Goriaev}}, \bibinfo {author} {\bibfnamefont
  {D.}~\bibnamefont {Gradic}}, \bibinfo {author} {\bibfnamefont
  {M.}~\bibnamefont {Grahl}}, \bibinfo {author} {\bibfnamefont
  {J.}~\bibnamefont {Green}}, \bibinfo {author} {\bibfnamefont
  {H.}~\bibnamefont {Greuner}}, \bibinfo {author} {\bibfnamefont
  {A.}~\bibnamefont {Grosman}}, \bibinfo {author} {\bibfnamefont
  {H.}~\bibnamefont {Grote}}, \bibinfo {author} {\bibfnamefont
  {M.}~\bibnamefont {Gruca}}, \bibinfo {author} {\bibfnamefont
  {O.}~\bibnamefont {Grulke}}, \bibinfo {author} {\bibfnamefont
  {C.}~\bibnamefont {Guerard}}, \bibinfo {author} {\bibfnamefont
  {P.}~\bibnamefont {Hacker}}, \bibinfo {author} {\bibfnamefont
  {X.}~\bibnamefont {Han}}, \bibinfo {author} {\bibfnamefont {J.~H.}\
  \bibnamefont {Harris}}, \bibinfo {author} {\bibfnamefont {D.}~\bibnamefont
  {Hartmann}}, \bibinfo {author} {\bibfnamefont {D.}~\bibnamefont
  {Hathiramani}}, \bibinfo {author} {\bibfnamefont {B.}~\bibnamefont {Hein}},
  \bibinfo {author} {\bibfnamefont {B.}~\bibnamefont {Heinemann}}, \bibinfo
  {author} {\bibfnamefont {P.}~\bibnamefont {Helander}}, \bibinfo {author}
  {\bibfnamefont {S.}~\bibnamefont {Henneberg}}, \bibinfo {author}
  {\bibfnamefont {M.}~\bibnamefont {Henkel}}, \bibinfo {author} {\bibfnamefont
  {U.}~\bibnamefont {Hergenhahn}}, \bibinfo {author} {\bibfnamefont {J.~H.}\
  \bibnamefont {Sanchez}}, \bibinfo {author} {\bibfnamefont {C.}~\bibnamefont
  {Hidalgo}}, \bibinfo {author} {\bibfnamefont {K.~P.}\ \bibnamefont
  {Hollfeld}}, \bibinfo {author} {\bibfnamefont {A.}~\bibnamefont
  {H\"{o}lting}}, \bibinfo {author} {\bibfnamefont {D.}~\bibnamefont
  {H\"{o}schen}}, \bibinfo {author} {\bibfnamefont {M.}~\bibnamefont {Houry}},
  \bibinfo {author} {\bibfnamefont {J.}~\bibnamefont {Howard}}, \bibinfo
  {author} {\bibfnamefont {X.}~\bibnamefont {Huang}}, \bibinfo {author}
  {\bibfnamefont {Z.}~\bibnamefont {Huang}}, \bibinfo {author} {\bibfnamefont
  {M.}~\bibnamefont {Hubeny}}, \bibinfo {author} {\bibfnamefont
  {M.}~\bibnamefont {Huber}}, \bibinfo {author} {\bibfnamefont
  {H.}~\bibnamefont {Hunger}}, \bibinfo {author} {\bibfnamefont
  {K.}~\bibnamefont {Ida}}, \bibinfo {author} {\bibfnamefont {T.}~\bibnamefont
  {Ilkei}}, \bibinfo {author} {\bibfnamefont {S.}~\bibnamefont {Illy}},
  \bibinfo {author} {\bibfnamefont {B.}~\bibnamefont {Israeli}}, \bibinfo
  {author} {\bibfnamefont {S.}~\bibnamefont {Jablonski}}, \bibinfo {author}
  {\bibfnamefont {M.}~\bibnamefont {Jakubowski}}, \bibinfo {author}
  {\bibfnamefont {J.}~\bibnamefont {Jelonnek}}, \bibinfo {author}
  {\bibfnamefont {H.}~\bibnamefont {Jenzsch}}, \bibinfo {author} {\bibfnamefont
  {T.}~\bibnamefont {Jesche}}, \bibinfo {author} {\bibfnamefont
  {M.}~\bibnamefont {Jia}}, \bibinfo {author} {\bibfnamefont {P.}~\bibnamefont
  {Junghanns}}, \bibinfo {author} {\bibfnamefont {J.}~\bibnamefont
  {Kacmarczyk}}, \bibinfo {author} {\bibfnamefont {J.-P.}\ \bibnamefont
  {Kallmeyer}}, \bibinfo {author} {\bibfnamefont {U.}~\bibnamefont {Kamionka}},
  \bibinfo {author} {\bibfnamefont {H.}~\bibnamefont {Kasahara}}, \bibinfo
  {author} {\bibfnamefont {W.}~\bibnamefont {Kasparek}}, \bibinfo {author}
  {\bibfnamefont {N.}~\bibnamefont {Kenmochi}}, \bibinfo {author}
  {\bibfnamefont {C.}~\bibnamefont {Killer}}, \bibinfo {author} {\bibfnamefont
  {A.}~\bibnamefont {Kirschner}}, \bibinfo {author} {\bibfnamefont
  {T.}~\bibnamefont {Klinger}}, \bibinfo {author} {\bibfnamefont
  {J.}~\bibnamefont {Knauer}}, \bibinfo {author} {\bibfnamefont
  {M.}~\bibnamefont {Knaup}}, \bibinfo {author} {\bibfnamefont
  {A.}~\bibnamefont {Knieps}}, \bibinfo {author} {\bibfnamefont
  {T.}~\bibnamefont {Kobarg}}, \bibinfo {author} {\bibfnamefont
  {G.}~\bibnamefont {Kocsis}}, \bibinfo {author} {\bibfnamefont
  {F.}~\bibnamefont {K\"{o}chl}}, \bibinfo {author} {\bibfnamefont
  {Y.}~\bibnamefont {Kolesnichenko}}, \bibinfo {author} {\bibfnamefont
  {A.}~\bibnamefont {K\"{o}nies}}, \bibinfo {author} {\bibfnamefont
  {R.}~\bibnamefont {K\"{o}nig}}, \bibinfo {author} {\bibfnamefont
  {P.}~\bibnamefont {Kornejew}}, \bibinfo {author} {\bibfnamefont {J.-P.}\
  \bibnamefont {Koschinsky}}, \bibinfo {author} {\bibfnamefont
  {F.}~\bibnamefont {K\"{o}ster}}, \bibinfo {author} {\bibfnamefont
  {M.}~\bibnamefont {Kr\"{a}mer}}, \bibinfo {author} {\bibfnamefont
  {R.}~\bibnamefont {Krampitz}}, \bibinfo {author} {\bibfnamefont
  {A.}~\bibnamefont {Kr\"{a}mer-Flecken}}, \bibinfo {author} {\bibfnamefont
  {N.}~\bibnamefont {Krawczyk}}, \bibinfo {author} {\bibfnamefont
  {T.}~\bibnamefont {Kremeyer}}, \bibinfo {author} {\bibfnamefont
  {J.}~\bibnamefont {Krom}}, \bibinfo {author} {\bibfnamefont {I.}~\bibnamefont
  {Ksiazek}}, \bibinfo {author} {\bibfnamefont {M.}~\bibnamefont {Kubkowska}},
  \bibinfo {author} {\bibfnamefont {G.}~\bibnamefont {K\"{u}hner}}, \bibinfo
  {author} {\bibfnamefont {T.}~\bibnamefont {Kurki-Suonio}}, \bibinfo {author}
  {\bibfnamefont {P.~A.}\ \bibnamefont {Kurz}}, \bibinfo {author}
  {\bibfnamefont {M.}~\bibnamefont {Landreman}}, \bibinfo {author}
  {\bibfnamefont {P.}~\bibnamefont {Lang}}, \bibinfo {author} {\bibfnamefont
  {R.}~\bibnamefont {Lang}}, \bibinfo {author} {\bibfnamefont {S.}~\bibnamefont
  {Langish}}, \bibinfo {author} {\bibfnamefont {H.}~\bibnamefont {Laqua}},
  \bibinfo {author} {\bibfnamefont {R.}~\bibnamefont {Laube}}, \bibinfo
  {author} {\bibfnamefont {S.}~\bibnamefont {Lazerson}}, \bibinfo {author}
  {\bibfnamefont {C.}~\bibnamefont {Lechte}}, \bibinfo {author} {\bibfnamefont
  {M.}~\bibnamefont {Lennartz}}, \bibinfo {author} {\bibfnamefont
  {W.}~\bibnamefont {Leonhardt}}, \bibinfo {author} {\bibfnamefont
  {C.}~\bibnamefont {Li}}, \bibinfo {author} {\bibfnamefont {C.}~\bibnamefont
  {Li}}, \bibinfo {author} {\bibfnamefont {Y.}~\bibnamefont {Li}}, \bibinfo
  {author} {\bibfnamefont {Y.}~\bibnamefont {Liang}}, \bibinfo {author}
  {\bibfnamefont {C.}~\bibnamefont {Linsmeier}}, \bibinfo {author}
  {\bibfnamefont {S.}~\bibnamefont {Liu}}, \bibinfo {author} {\bibfnamefont
  {J.-F.}\ \bibnamefont {Lobsien}}, \bibinfo {author} {\bibfnamefont
  {D.}~\bibnamefont {Loesser}}, \bibinfo {author} {\bibfnamefont {J.~L.}\
  \bibnamefont {Cisquella}}, \bibinfo {author} {\bibfnamefont {J.}~\bibnamefont
  {Lore}}, \bibinfo {author} {\bibfnamefont {A.}~\bibnamefont {Lorenz}},
  \bibinfo {author} {\bibfnamefont {M.}~\bibnamefont {Losert}}, \bibinfo
  {author} {\bibfnamefont {A.}~\bibnamefont {L\"{u}cke}}, \bibinfo {author}
  {\bibfnamefont {A.}~\bibnamefont {Lumsdaine}}, \bibinfo {author}
  {\bibfnamefont {V.}~\bibnamefont {Lutsenko}}, \bibinfo {author}
  {\bibfnamefont {H.}~\bibnamefont {Maa{\ss}berg}}, \bibinfo {author}
  {\bibfnamefont {O.}~\bibnamefont {Marchuk}}, \bibinfo {author} {\bibfnamefont
  {J.~H.}\ \bibnamefont {Matthew}}, \bibinfo {author} {\bibfnamefont
  {S.}~\bibnamefont {Marsen}}, \bibinfo {author} {\bibfnamefont
  {M.}~\bibnamefont {Marushchenko}}, \bibinfo {author} {\bibfnamefont
  {S.}~\bibnamefont {Masuzaki}}, \bibinfo {author} {\bibfnamefont
  {D.}~\bibnamefont {Maurer}}, \bibinfo {author} {\bibfnamefont
  {M.}~\bibnamefont {Mayer}}, \bibinfo {author} {\bibfnamefont
  {K.}~\bibnamefont {McCarthy}}, \bibinfo {author} {\bibfnamefont
  {P.}~\bibnamefont {McNeely}}, \bibinfo {author} {\bibfnamefont
  {A.}~\bibnamefont {Meier}}, \bibinfo {author} {\bibfnamefont
  {D.}~\bibnamefont {Mellein}}, \bibinfo {author} {\bibfnamefont
  {B.}~\bibnamefont {Mendelevitch}}, \bibinfo {author} {\bibfnamefont
  {P.}~\bibnamefont {Mertens}}, \bibinfo {author} {\bibfnamefont
  {D.}~\bibnamefont {Mikkelsen}}, \bibinfo {author} {\bibfnamefont
  {A.}~\bibnamefont {Mishchenko}}, \bibinfo {author} {\bibfnamefont
  {B.}~\bibnamefont {Missal}}, \bibinfo {author} {\bibfnamefont
  {J.}~\bibnamefont {Mittelstaedt}}, \bibinfo {author} {\bibfnamefont
  {T.}~\bibnamefont {Mizuuchi}}, \bibinfo {author} {\bibfnamefont
  {A.}~\bibnamefont {Mollen}}, \bibinfo {author} {\bibfnamefont
  {V.}~\bibnamefont {Moncada}}, \bibinfo {author} {\bibfnamefont
  {T.}~\bibnamefont {M\"{o}nnich}}, \bibinfo {author} {\bibfnamefont
  {T.}~\bibnamefont {Morisaki}}, \bibinfo {author} {\bibfnamefont
  {D.}~\bibnamefont {Moseev}}, \bibinfo {author} {\bibfnamefont
  {S.}~\bibnamefont {Murakami}}, \bibinfo {author} {\bibfnamefont
  {G.}~\bibnamefont {N{\'{a}}fr{\'{a}}di}}, \bibinfo {author} {\bibfnamefont
  {M.}~\bibnamefont {Nagel}}, \bibinfo {author} {\bibfnamefont
  {D.}~\bibnamefont {Naujoks}}, \bibinfo {author} {\bibfnamefont
  {H.}~\bibnamefont {Neilson}}, \bibinfo {author} {\bibfnamefont
  {R.}~\bibnamefont {Neu}}, \bibinfo {author} {\bibfnamefont {O.}~\bibnamefont
  {Neubauer}}, \bibinfo {author} {\bibfnamefont {T.}~\bibnamefont {Ngo}},
  \bibinfo {author} {\bibfnamefont {D.}~\bibnamefont {Nicolai}}, \bibinfo
  {author} {\bibfnamefont {S.~K.}\ \bibnamefont {Nielsen}}, \bibinfo {author}
  {\bibfnamefont {H.}~\bibnamefont {Niemann}}, \bibinfo {author} {\bibfnamefont
  {T.}~\bibnamefont {Nishizawa}}, \bibinfo {author} {\bibfnamefont
  {R.}~\bibnamefont {Nocentini}}, \bibinfo {author} {\bibfnamefont
  {C.}~\bibnamefont {N\"{u}hrenberg}}, \bibinfo {author} {\bibfnamefont
  {J.}~\bibnamefont {N\"{u}hrenberg}}, \bibinfo {author} {\bibfnamefont
  {S.}~\bibnamefont {Obermayer}}, \bibinfo {author} {\bibfnamefont
  {G.}~\bibnamefont {Offermanns}}, \bibinfo {author} {\bibfnamefont
  {K.}~\bibnamefont {Ogawa}}, \bibinfo {author} {\bibfnamefont
  {J.}~\bibnamefont {\"{O}lmanns}}, \bibinfo {author} {\bibfnamefont
  {J.}~\bibnamefont {Ongena}}, \bibinfo {author} {\bibfnamefont {J.~W.}\
  \bibnamefont {Oosterbeek}}, \bibinfo {author} {\bibfnamefont
  {G.}~\bibnamefont {Orozco}}, \bibinfo {author} {\bibfnamefont
  {M.}~\bibnamefont {Otte}}, \bibinfo {author} {\bibfnamefont {L.~P.}\
  \bibnamefont {Rodriguez}}, \bibinfo {author} {\bibfnamefont {N.}~\bibnamefont
  {Panadero}}, \bibinfo {author} {\bibfnamefont {N.~P.}\ \bibnamefont
  {Alvarez}}, \bibinfo {author} {\bibfnamefont {D.}~\bibnamefont
  {Papenfu{\ss}}}, \bibinfo {author} {\bibfnamefont {S.}~\bibnamefont {Paqay}},
  \bibinfo {author} {\bibfnamefont {E.}~\bibnamefont {Pawelec}}, \bibinfo
  {author} {\bibfnamefont {T.~S.}\ \bibnamefont {Pedersen}}, \bibinfo {author}
  {\bibfnamefont {G.}~\bibnamefont {Pelka}}, \bibinfo {author} {\bibfnamefont
  {V.}~\bibnamefont {Perseo}}, \bibinfo {author} {\bibfnamefont
  {B.}~\bibnamefont {Peterson}}, \bibinfo {author} {\bibfnamefont
  {D.}~\bibnamefont {Pilopp}}, \bibinfo {author} {\bibfnamefont
  {S.}~\bibnamefont {Pingel}}, \bibinfo {author} {\bibfnamefont
  {F.}~\bibnamefont {Pisano}}, \bibinfo {author} {\bibfnamefont
  {B.}~\bibnamefont {Plaum}}, \bibinfo {author} {\bibfnamefont
  {G.}~\bibnamefont {Plunk}}, \bibinfo {author} {\bibfnamefont
  {P.}~\bibnamefont {P\"{o}l\"{o}skei}}, \bibinfo {author} {\bibfnamefont
  {M.}~\bibnamefont {Porkolab}}, \bibinfo {author} {\bibfnamefont
  {J.}~\bibnamefont {Proll}}, \bibinfo {author} {\bibfnamefont {M.-E.}\
  \bibnamefont {Puiatti}}, \bibinfo {author} {\bibfnamefont {A.~P.}\
  \bibnamefont {Sitjes}}, \bibinfo {author} {\bibfnamefont {F.}~\bibnamefont
  {Purps}}, \bibinfo {author} {\bibfnamefont {M.}~\bibnamefont {Rack}},
  \bibinfo {author} {\bibfnamefont {S.}~\bibnamefont {R{\'{e}}csei}}, \bibinfo
  {author} {\bibfnamefont {A.}~\bibnamefont {Reiman}}, \bibinfo {author}
  {\bibfnamefont {F.}~\bibnamefont {Reimold}}, \bibinfo {author} {\bibfnamefont
  {D.}~\bibnamefont {Reiter}}, \bibinfo {author} {\bibfnamefont
  {F.}~\bibnamefont {Remppel}}, \bibinfo {author} {\bibfnamefont
  {S.}~\bibnamefont {Renard}}, \bibinfo {author} {\bibfnamefont
  {R.}~\bibnamefont {Riedl}}, \bibinfo {author} {\bibfnamefont
  {J.}~\bibnamefont {Riemann}}, \bibinfo {author} {\bibfnamefont
  {K.}~\bibnamefont {Risse}}, \bibinfo {author} {\bibfnamefont
  {V.}~\bibnamefont {Rohde}}, \bibinfo {author} {\bibfnamefont
  {H.}~\bibnamefont {R\"{o}hlinger}}, \bibinfo {author} {\bibfnamefont
  {M.}~\bibnamefont {Rom{\'{e}}}}, \bibinfo {author} {\bibfnamefont
  {D.}~\bibnamefont {Rondeshagen}}, \bibinfo {author} {\bibfnamefont
  {P.}~\bibnamefont {Rong}}, \bibinfo {author} {\bibfnamefont {B.}~\bibnamefont
  {Roth}}, \bibinfo {author} {\bibfnamefont {L.}~\bibnamefont {Rudischhauser}},
  \bibinfo {author} {\bibfnamefont {K.}~\bibnamefont {Rummel}}, \bibinfo
  {author} {\bibfnamefont {T.}~\bibnamefont {Rummel}}, \bibinfo {author}
  {\bibfnamefont {A.}~\bibnamefont {Runov}}, \bibinfo {author} {\bibfnamefont
  {N.}~\bibnamefont {Rust}}, \bibinfo {author} {\bibfnamefont {L.}~\bibnamefont
  {Ryc}}, \bibinfo {author} {\bibfnamefont {S.}~\bibnamefont {Ryosuke}},
  \bibinfo {author} {\bibfnamefont {R.}~\bibnamefont {Sakamoto}}, \bibinfo
  {author} {\bibfnamefont {M.}~\bibnamefont {Salewski}}, \bibinfo {author}
  {\bibfnamefont {A.}~\bibnamefont {Samartsev}}, \bibinfo {author}
  {\bibfnamefont {E.}~\bibnamefont {S{\'{a}}nchez}}, \bibinfo {author}
  {\bibfnamefont {F.}~\bibnamefont {Sano}}, \bibinfo {author} {\bibfnamefont
  {S.}~\bibnamefont {Satake}}, \bibinfo {author} {\bibfnamefont
  {J.}~\bibnamefont {Schacht}}, \bibinfo {author} {\bibfnamefont
  {G.}~\bibnamefont {Satheeswaran}}, \bibinfo {author} {\bibfnamefont
  {F.}~\bibnamefont {Schauer}}, \bibinfo {author} {\bibfnamefont
  {T.}~\bibnamefont {Scherer}}, \bibinfo {author} {\bibfnamefont
  {A.}~\bibnamefont {Schlaich}}, \bibinfo {author} {\bibfnamefont
  {G.}~\bibnamefont {Schlisio}}, \bibinfo {author} {\bibfnamefont
  {F.}~\bibnamefont {Schluck}}, \bibinfo {author} {\bibfnamefont {K.-H.}\
  \bibnamefont {Schl\"{u}ter}}, \bibinfo {author} {\bibfnamefont
  {J.}~\bibnamefont {Schmitt}}, \bibinfo {author} {\bibfnamefont
  {H.}~\bibnamefont {Schmitz}}, \bibinfo {author} {\bibfnamefont
  {O.}~\bibnamefont {Schmitz}}, \bibinfo {author} {\bibfnamefont
  {S.}~\bibnamefont {Schmuck}}, \bibinfo {author} {\bibfnamefont
  {M.}~\bibnamefont {Schneider}}, \bibinfo {author} {\bibfnamefont
  {W.}~\bibnamefont {Schneider}}, \bibinfo {author} {\bibfnamefont
  {P.}~\bibnamefont {Scholz}}, \bibinfo {author} {\bibfnamefont
  {R.}~\bibnamefont {Schrittwieser}}, \bibinfo {author} {\bibfnamefont
  {M.}~\bibnamefont {Schr\"{o}der}}, \bibinfo {author} {\bibfnamefont
  {T.}~\bibnamefont {Schr\"{o}der}}, \bibinfo {author} {\bibfnamefont
  {R.}~\bibnamefont {Schroeder}}, \bibinfo {author} {\bibfnamefont
  {H.}~\bibnamefont {Schumacher}}, \bibinfo {author} {\bibfnamefont
  {B.}~\bibnamefont {Schweer}}, \bibinfo {author} {\bibfnamefont
  {S.}~\bibnamefont {Sereda}}, \bibinfo {author} {\bibfnamefont
  {B.}~\bibnamefont {Shanahan}}, \bibinfo {author} {\bibfnamefont
  {M.}~\bibnamefont {Sibilia}}, \bibinfo {author} {\bibfnamefont
  {P.}~\bibnamefont {Sinha}}, \bibinfo {author} {\bibfnamefont
  {S.}~\bibnamefont {Sipli\"{a}}}, \bibinfo {author} {\bibfnamefont
  {C.}~\bibnamefont {Slaby}}, \bibinfo {author} {\bibfnamefont
  {M.}~\bibnamefont {Sleczka}}, \bibinfo {author} {\bibfnamefont
  {W.}~\bibnamefont {Spiess}}, \bibinfo {author} {\bibfnamefont {D.~A.}\
  \bibnamefont {Spong}}, \bibinfo {author} {\bibfnamefont {A.}~\bibnamefont
  {Spring}}, \bibinfo {author} {\bibfnamefont {R.}~\bibnamefont {Stadler}},
  \bibinfo {author} {\bibfnamefont {M.}~\bibnamefont {Stejner}}, \bibinfo
  {author} {\bibfnamefont {L.}~\bibnamefont {Stephey}}, \bibinfo {author}
  {\bibfnamefont {U.}~\bibnamefont {Stridde}}, \bibinfo {author} {\bibfnamefont
  {C.}~\bibnamefont {Suzuki}}, \bibinfo {author} {\bibfnamefont
  {V.}~\bibnamefont {Szab{\'{o}}}}, \bibinfo {author} {\bibfnamefont
  {T.}~\bibnamefont {Szabolics}}, \bibinfo {author} {\bibfnamefont
  {T.}~\bibnamefont {Szepesi}}, \bibinfo {author} {\bibfnamefont
  {Z.}~\bibnamefont {Sz\"{o}kefalvi-Nagy}}, \bibinfo {author} {\bibfnamefont
  {N.}~\bibnamefont {Tamura}}, \bibinfo {author} {\bibfnamefont
  {A.}~\bibnamefont {Tancetti}}, \bibinfo {author} {\bibfnamefont
  {J.}~\bibnamefont {Terry}}, \bibinfo {author} {\bibfnamefont
  {J.}~\bibnamefont {Thomas}}, \bibinfo {author} {\bibfnamefont
  {M.}~\bibnamefont {Thumm}}, \bibinfo {author} {\bibfnamefont {J.~M.}\
  \bibnamefont {Travere}}, \bibinfo {author} {\bibfnamefont {P.}~\bibnamefont
  {Traverso}}, \bibinfo {author} {\bibfnamefont {J.}~\bibnamefont {Tretter}},
  \bibinfo {author} {\bibfnamefont {H.~T.}\ \bibnamefont {Mora}}, \bibinfo
  {author} {\bibfnamefont {H.}~\bibnamefont {Tsuchiya}}, \bibinfo {author}
  {\bibfnamefont {T.}~\bibnamefont {Tsujimura}}, \bibinfo {author}
  {\bibfnamefont {S.}~\bibnamefont {Tulip{\'{a}}n}}, \bibinfo {author}
  {\bibfnamefont {B.}~\bibnamefont {Unterberg}}, \bibinfo {author}
  {\bibfnamefont {I.}~\bibnamefont {Vakulchyk}}, \bibinfo {author}
  {\bibfnamefont {S.}~\bibnamefont {Valet}}, \bibinfo {author} {\bibfnamefont
  {L.}~\bibnamefont {Van{\'{o}}}}, \bibinfo {author} {\bibfnamefont
  {P.}~\bibnamefont {van Eeten}}, \bibinfo {author} {\bibfnamefont
  {B.}~\bibnamefont {van Milligen}}, \bibinfo {author} {\bibfnamefont {A.~J.}\
  \bibnamefont {van Vuuren}}, \bibinfo {author} {\bibfnamefont
  {L.}~\bibnamefont {Vela}}, \bibinfo {author} {\bibfnamefont {J.-L.}\
  \bibnamefont {Velasco}}, \bibinfo {author} {\bibfnamefont {M.}~\bibnamefont
  {Vergote}}, \bibinfo {author} {\bibfnamefont {M.}~\bibnamefont {Vervier}},
  \bibinfo {author} {\bibfnamefont {N.}~\bibnamefont {Vianello}}, \bibinfo
  {author} {\bibfnamefont {H.}~\bibnamefont {Viebke}}, \bibinfo {author}
  {\bibfnamefont {R.}~\bibnamefont {Vilbrandt}}, \bibinfo {author}
  {\bibfnamefont {A.}~\bibnamefont {von Stechow}}, \bibinfo {author}
  {\bibfnamefont {A.}~\bibnamefont {Vork\"{o}per}}, \bibinfo {author}
  {\bibfnamefont {S.}~\bibnamefont {Wadle}}, \bibinfo {author} {\bibfnamefont
  {F.}~\bibnamefont {Wagner}}, \bibinfo {author} {\bibfnamefont
  {E.}~\bibnamefont {Wang}}, \bibinfo {author} {\bibfnamefont {N.}~\bibnamefont
  {Wang}}, \bibinfo {author} {\bibfnamefont {Z.}~\bibnamefont {Wang}}, \bibinfo
  {author} {\bibfnamefont {T.}~\bibnamefont {Wauters}}, \bibinfo {author}
  {\bibfnamefont {L.}~\bibnamefont {Wegener}}, \bibinfo {author} {\bibfnamefont
  {J.}~\bibnamefont {Weggen}}, \bibinfo {author} {\bibfnamefont
  {T.}~\bibnamefont {Wegner}}, \bibinfo {author} {\bibfnamefont
  {Y.}~\bibnamefont {Wei}}, \bibinfo {author} {\bibfnamefont {G.}~\bibnamefont
  {Weir}}, \bibinfo {author} {\bibfnamefont {J.}~\bibnamefont {Wendorf}},
  \bibinfo {author} {\bibfnamefont {U.}~\bibnamefont {Wenzel}}, \bibinfo
  {author} {\bibfnamefont {A.}~\bibnamefont {Werner}}, \bibinfo {author}
  {\bibfnamefont {A.}~\bibnamefont {White}}, \bibinfo {author} {\bibfnamefont
  {B.}~\bibnamefont {Wiegel}}, \bibinfo {author} {\bibfnamefont
  {F.}~\bibnamefont {Wilde}}, \bibinfo {author} {\bibfnamefont
  {T.}~\bibnamefont {Windisch}}, \bibinfo {author} {\bibfnamefont
  {M.}~\bibnamefont {Winkler}}, \bibinfo {author} {\bibfnamefont
  {A.}~\bibnamefont {Winter}}, \bibinfo {author} {\bibfnamefont
  {V.}~\bibnamefont {Winters}}, \bibinfo {author} {\bibfnamefont
  {S.}~\bibnamefont {Wolf}}, \bibinfo {author} {\bibfnamefont {R.~C.}\
  \bibnamefont {Wolf}}, \bibinfo {author} {\bibfnamefont {A.}~\bibnamefont
  {Wright}}, \bibinfo {author} {\bibfnamefont {G.}~\bibnamefont {Wurden}},
  \bibinfo {author} {\bibfnamefont {P.}~\bibnamefont {Xanthopoulos}}, \bibinfo
  {author} {\bibfnamefont {H.}~\bibnamefont {Yamada}}, \bibinfo {author}
  {\bibfnamefont {I.}~\bibnamefont {Yamada}}, \bibinfo {author} {\bibfnamefont
  {R.}~\bibnamefont {Yasuhara}}, \bibinfo {author} {\bibfnamefont
  {M.}~\bibnamefont {Yokoyama}}, \bibinfo {author} {\bibfnamefont
  {M.}~\bibnamefont {Zanini}}, \bibinfo {author} {\bibfnamefont
  {M.}~\bibnamefont {Zarnstorff}}, \bibinfo {author} {\bibfnamefont
  {A.}~\bibnamefont {Zeitler}}, \bibinfo {author} {\bibfnamefont
  {H.}~\bibnamefont {Zhang}}, \bibinfo {author} {\bibfnamefont
  {J.}~\bibnamefont {Zhu}}, \bibinfo {author} {\bibfnamefont {M.}~\bibnamefont
  {Zilker}}, \bibinfo {author} {\bibfnamefont {A.}~\bibnamefont {Zocco}},
  \bibinfo {author} {\bibfnamefont {S.}~\bibnamefont {Zoletnik}}, \ and\
  \bibinfo {author} {\bibfnamefont {M.~Z.}\ \bibnamefont {and}},\ }\href
  {\doibase 10.1038/s41586-021-04023-y} {\bibfield  {journal} {\bibinfo
  {journal} {Nature}\ }\textbf {\bibinfo {volume} {598}},\ \bibinfo {pages}
  {E5} (\bibinfo {year} {2021})}\BibitemShut {NoStop}%
\bibitem [{\citenamefont {Dinklage}\ \emph {et~al.}(2013)\citenamefont
  {Dinklage}, \citenamefont {Yokoyama}, \citenamefont {Tanaka}, \citenamefont
  {Velasco}, \citenamefont {L\'opez-Bruna}, \citenamefont {Beidler},
  \citenamefont {Satake}, \citenamefont {Ascas\'ibar}, \citenamefont
  {Ar\'evalo}, \citenamefont {Baldzuhn}, \citenamefont {Feng}, \citenamefont
  {Gates}, \citenamefont {Geiger}, \citenamefont {Ida}, \citenamefont
  {Jakubowski}, \citenamefont {L\'opez-Fraguas}, \citenamefont {Maa{\ss}berg},
  \citenamefont {Miyazawa}, \citenamefont {Morisaki}, \citenamefont {Murakami},
  \citenamefont {Pablant}, \citenamefont {Kobayashi}, \citenamefont {Seki},
  \citenamefont {Suzuki}, \citenamefont {Suzuki}, \citenamefont {Turkin},
  \citenamefont {Wakasa}, \citenamefont {Wolf}, \citenamefont {Yamada},
  \citenamefont {Yoshinuma}, \citenamefont {{LHD Exp. Group}}, \citenamefont
  {{TJ-II Team}},\ and\ \citenamefont {{W7-AS Team}}}]{Dinklage_nf_2013}%
  \BibitemOpen
  \bibfield  {author} {\bibinfo {author} {\bibfnamefont {A.}~\bibnamefont
  {Dinklage}}, \bibinfo {author} {\bibfnamefont {M.}~\bibnamefont {Yokoyama}},
  \bibinfo {author} {\bibfnamefont {K.}~\bibnamefont {Tanaka}}, \bibinfo
  {author} {\bibfnamefont {J.~L.}\ \bibnamefont {Velasco}}, \bibinfo {author}
  {\bibfnamefont {D.}~\bibnamefont {L\'opez-Bruna}}, \bibinfo {author}
  {\bibfnamefont {C.~D.}\ \bibnamefont {Beidler}}, \bibinfo {author}
  {\bibfnamefont {S.}~\bibnamefont {Satake}}, \bibinfo {author} {\bibfnamefont
  {E.}~\bibnamefont {Ascas\'ibar}}, \bibinfo {author} {\bibfnamefont
  {J.}~\bibnamefont {Ar\'evalo}}, \bibinfo {author} {\bibfnamefont
  {J.}~\bibnamefont {Baldzuhn}}, \bibinfo {author} {\bibfnamefont
  {Y.}~\bibnamefont {Feng}}, \bibinfo {author} {\bibfnamefont {D.}~\bibnamefont
  {Gates}}, \bibinfo {author} {\bibfnamefont {J.}~\bibnamefont {Geiger}},
  \bibinfo {author} {\bibfnamefont {K.}~\bibnamefont {Ida}}, \bibinfo {author}
  {\bibfnamefont {M.}~\bibnamefont {Jakubowski}}, \bibinfo {author}
  {\bibfnamefont {A.}~\bibnamefont {L\'opez-Fraguas}}, \bibinfo {author}
  {\bibfnamefont {H.}~\bibnamefont {Maa{\ss}berg}}, \bibinfo {author}
  {\bibfnamefont {J.}~\bibnamefont {Miyazawa}}, \bibinfo {author}
  {\bibfnamefont {T.}~\bibnamefont {Morisaki}}, \bibinfo {author}
  {\bibfnamefont {S.}~\bibnamefont {Murakami}}, \bibinfo {author}
  {\bibfnamefont {N.}~\bibnamefont {Pablant}}, \bibinfo {author} {\bibfnamefont
  {S.}~\bibnamefont {Kobayashi}}, \bibinfo {author} {\bibfnamefont
  {R.}~\bibnamefont {Seki}}, \bibinfo {author} {\bibfnamefont {C.}~\bibnamefont
  {Suzuki}}, \bibinfo {author} {\bibfnamefont {Y.}~\bibnamefont {Suzuki}},
  \bibinfo {author} {\bibfnamefont {Y.}~\bibnamefont {Turkin}}, \bibinfo
  {author} {\bibfnamefont {A.}~\bibnamefont {Wakasa}}, \bibinfo {author}
  {\bibfnamefont {R.}~\bibnamefont {Wolf}}, \bibinfo {author} {\bibfnamefont
  {H.}~\bibnamefont {Yamada}}, \bibinfo {author} {\bibfnamefont
  {M.}~\bibnamefont {Yoshinuma}}, \bibinfo {author} {\bibnamefont {{LHD Exp.
  Group}}}, \bibinfo {author} {\bibnamefont {{TJ-II Team}}}, \ and\ \bibinfo
  {author} {\bibnamefont {{W7-AS Team}}},\ }\href
  {http://stacks.iop.org/0029-5515/53/i=6/a=063022} {\bibfield  {journal}
  {\bibinfo  {journal} {Nuclear Fusion}\ }\textbf {\bibinfo {volume} {53}},\
  \bibinfo {pages} {063022} (\bibinfo {year} {2013})}\BibitemShut {NoStop}%
\bibitem [{\citenamefont {Geiger}\ \emph {et~al.}(2014)\citenamefont {Geiger},
  \citenamefont {Beidler}, \citenamefont {Feng}, \citenamefont {Maa{\ss}berg},
  \citenamefont {Marushchenko},\ and\ \citenamefont {Turkin}}]{Geiger_2015}%
  \BibitemOpen
  \bibfield  {author} {\bibinfo {author} {\bibfnamefont {J.}~\bibnamefont
  {Geiger}}, \bibinfo {author} {\bibfnamefont {C.~D.}\ \bibnamefont {Beidler}},
  \bibinfo {author} {\bibfnamefont {Y.}~\bibnamefont {Feng}}, \bibinfo {author}
  {\bibfnamefont {H.}~\bibnamefont {Maa{\ss}berg}}, \bibinfo {author}
  {\bibfnamefont {N.~B.}\ \bibnamefont {Marushchenko}}, \ and\ \bibinfo
  {author} {\bibfnamefont {Y.}~\bibnamefont {Turkin}},\ }\href {\doibase
  10.1088/0741-3335/57/1/014004} {\bibfield  {journal} {\bibinfo  {journal}
  {Plasma Physics and Controlled Fusion}\ }\textbf {\bibinfo {volume} {57}},\
  \bibinfo {pages} {014004} (\bibinfo {year} {2014})}\BibitemShut {NoStop}%
\bibitem [{\citenamefont {Maa{\ss}berg}\ \emph {et~al.}(1999)\citenamefont
  {Maa{\ss}berg}, \citenamefont {Beidler},\ and\ \citenamefont
  {Simmet}}]{Maassberg_ppcf_1999}%
  \BibitemOpen
  \bibfield  {author} {\bibinfo {author} {\bibfnamefont {H.}~\bibnamefont
  {Maa{\ss}berg}}, \bibinfo {author} {\bibfnamefont {C.~D.}\ \bibnamefont
  {Beidler}}, \ and\ \bibinfo {author} {\bibfnamefont {E.~E.}\ \bibnamefont
  {Simmet}},\ }\href {\doibase 10.1088/0741-3335/41/9/306} {\bibfield
  {journal} {\bibinfo  {journal} {Plasma Physics and Controlled Fusion}\
  }\textbf {\bibinfo {volume} {41}},\ \bibinfo {pages} {1135} (\bibinfo {year}
  {1999})}\BibitemShut {NoStop}%
\bibitem [{\citenamefont {Beidler}\ \emph {et~al.}(2018)\citenamefont
  {Beidler}, \citenamefont {Feng}, \citenamefont {Geiger}, \citenamefont
  {K\"{o}chl}, \citenamefont {Maa{\ss}berg}, \citenamefont {Marushchenko},
  \citenamefont {N\"{u}hrenberg}, \citenamefont {Smith},\ and\ \citenamefont
  {Turkin}}]{Beidler_ppcf_2018}%
  \BibitemOpen
  \bibfield  {author} {\bibinfo {author} {\bibfnamefont {C.~D.}\ \bibnamefont
  {Beidler}}, \bibinfo {author} {\bibfnamefont {Y.}~\bibnamefont {Feng}},
  \bibinfo {author} {\bibfnamefont {J.}~\bibnamefont {Geiger}}, \bibinfo
  {author} {\bibfnamefont {F.}~\bibnamefont {K\"{o}chl}}, \bibinfo {author}
  {\bibfnamefont {H.}~\bibnamefont {Maa{\ss}berg}}, \bibinfo {author}
  {\bibfnamefont {N.~B.}\ \bibnamefont {Marushchenko}}, \bibinfo {author}
  {\bibfnamefont {C.}~\bibnamefont {N\"{u}hrenberg}}, \bibinfo {author}
  {\bibfnamefont {H.~M.}\ \bibnamefont {Smith}}, \ and\ \bibinfo {author}
  {\bibfnamefont {Y.}~\bibnamefont {Turkin}},\ }\href {\doibase
  10.1088/1361-6587/aad970} {\bibfield  {journal} {\bibinfo  {journal} {Plasma
  Physics and Controlled Fusion}\ }\textbf {\bibinfo {volume} {60}},\ \bibinfo
  {pages} {105008} (\bibinfo {year} {2018})}\BibitemShut {NoStop}%
\bibitem [{\citenamefont {Velasco}\ \emph {et~al.}(2021)\citenamefont
  {Velasco}, \citenamefont {Calvo}, \citenamefont {Parra}, \citenamefont
  {d'Herbemont}, \citenamefont {Smith}, \citenamefont {Carralero},\ and\
  \citenamefont {Estrada}}]{Velasco_nf_2021}%
  \BibitemOpen
  \bibfield  {author} {\bibinfo {author} {\bibfnamefont {J.~L.}\ \bibnamefont
  {Velasco}}, \bibinfo {author} {\bibfnamefont {I.}~\bibnamefont {Calvo}},
  \bibinfo {author} {\bibfnamefont {F.~I.}\ \bibnamefont {Parra}}, \bibinfo
  {author} {\bibfnamefont {V.}~\bibnamefont {d'Herbemont}}, \bibinfo {author}
  {\bibfnamefont {H.~M.}\ \bibnamefont {Smith}}, \bibinfo {author}
  {\bibfnamefont {D.}~\bibnamefont {Carralero}}, \ and\ \bibinfo {author}
  {\bibfnamefont {T.}~\bibnamefont {Estrada}},\ }\href {\doibase
  10.1088/1741-4326/ac1dc7} {\bibfield  {journal} {\bibinfo  {journal} {Nuclear
  Fusion}\ }\textbf {\bibinfo {volume} {61}},\ \bibinfo {pages} {11603}
  (\bibinfo {year} {2021})}\BibitemShut {NoStop}%
\bibitem [{\citenamefont {Beurskens}\ \emph {et~al.}(2021)\citenamefont
  {Beurskens}, \citenamefont {Bozhenkov}, \citenamefont {Ford}, \citenamefont
  {Xanthopoulos}, \citenamefont {Zocco}, \citenamefont {Turkin}, \citenamefont
  {Alonso}, \citenamefont {Beidler}, \citenamefont {Calvo}, \citenamefont
  {Carralero}, \citenamefont {Estrada}, \citenamefont {Fuchert}, \citenamefont
  {Grulke}, \citenamefont {Hirsch}, \citenamefont {Ida}, \citenamefont
  {Jakubowski}, \citenamefont {Killer}, \citenamefont {Krychowiak},
  \citenamefont {Kwak}, \citenamefont {Lazerson}, \citenamefont {Langenberg},
  \citenamefont {Lunsford}, \citenamefont {Pablant}, \citenamefont {Pasch},
  \citenamefont {Pavone}, \citenamefont {Reimold}, \citenamefont {Romba},
  \citenamefont {von Stechow}, \citenamefont {Smith}, \citenamefont {Windisch},
  \citenamefont {Yoshinuma}, \citenamefont {Zhang}, \citenamefont {Wolf},\ and\
  \citenamefont {the W7-X~Team}}]{Beurskens_nf_2021}%
  \BibitemOpen
  \bibfield  {author} {\bibinfo {author} {\bibfnamefont {M.}~\bibnamefont
  {Beurskens}}, \bibinfo {author} {\bibfnamefont {S.}~\bibnamefont
  {Bozhenkov}}, \bibinfo {author} {\bibfnamefont {O.}~\bibnamefont {Ford}},
  \bibinfo {author} {\bibfnamefont {P.}~\bibnamefont {Xanthopoulos}}, \bibinfo
  {author} {\bibfnamefont {A.}~\bibnamefont {Zocco}}, \bibinfo {author}
  {\bibfnamefont {Y.}~\bibnamefont {Turkin}}, \bibinfo {author} {\bibfnamefont
  {A.}~\bibnamefont {Alonso}}, \bibinfo {author} {\bibfnamefont
  {C.}~\bibnamefont {Beidler}}, \bibinfo {author} {\bibfnamefont
  {I.}~\bibnamefont {Calvo}}, \bibinfo {author} {\bibfnamefont
  {D.}~\bibnamefont {Carralero}}, \bibinfo {author} {\bibfnamefont
  {T.}~\bibnamefont {Estrada}}, \bibinfo {author} {\bibfnamefont
  {G.}~\bibnamefont {Fuchert}}, \bibinfo {author} {\bibfnamefont
  {O.}~\bibnamefont {Grulke}}, \bibinfo {author} {\bibfnamefont
  {M.}~\bibnamefont {Hirsch}}, \bibinfo {author} {\bibfnamefont
  {K.}~\bibnamefont {Ida}}, \bibinfo {author} {\bibfnamefont {M.}~\bibnamefont
  {Jakubowski}}, \bibinfo {author} {\bibfnamefont {C.}~\bibnamefont {Killer}},
  \bibinfo {author} {\bibfnamefont {M.}~\bibnamefont {Krychowiak}}, \bibinfo
  {author} {\bibfnamefont {S.}~\bibnamefont {Kwak}}, \bibinfo {author}
  {\bibfnamefont {S.}~\bibnamefont {Lazerson}}, \bibinfo {author}
  {\bibfnamefont {A.}~\bibnamefont {Langenberg}}, \bibinfo {author}
  {\bibfnamefont {R.}~\bibnamefont {Lunsford}}, \bibinfo {author}
  {\bibfnamefont {N.}~\bibnamefont {Pablant}}, \bibinfo {author} {\bibfnamefont
  {E.}~\bibnamefont {Pasch}}, \bibinfo {author} {\bibfnamefont
  {A.}~\bibnamefont {Pavone}}, \bibinfo {author} {\bibfnamefont
  {F.}~\bibnamefont {Reimold}}, \bibinfo {author} {\bibfnamefont
  {T.}~\bibnamefont {Romba}}, \bibinfo {author} {\bibfnamefont
  {A.}~\bibnamefont {von Stechow}}, \bibinfo {author} {\bibfnamefont
  {H.}~\bibnamefont {Smith}}, \bibinfo {author} {\bibfnamefont
  {T.}~\bibnamefont {Windisch}}, \bibinfo {author} {\bibfnamefont
  {M.}~\bibnamefont {Yoshinuma}}, \bibinfo {author} {\bibfnamefont
  {D.}~\bibnamefont {Zhang}}, \bibinfo {author} {\bibfnamefont
  {R.}~\bibnamefont {Wolf}}, \ and\ \bibinfo {author} {\bibnamefont {the
  W7-X~Team}},\ }\href {\doibase 10.1088/1741-4326/ac1653} {\bibfield
  {journal} {\bibinfo  {journal} {Nuclear Fusion}\ }\textbf {\bibinfo {volume}
  {61}},\ \bibinfo {pages} {116072} (\bibinfo {year} {2021})}\BibitemShut
  {NoStop}%
\bibitem [{\citenamefont {Wolf}\ \emph {et~al.}(2017)\citenamefont {Wolf},
  \citenamefont {Ali}, \citenamefont {Alonso}, \citenamefont {Baldzuhn},
  \citenamefont {Beidler}, \citenamefont {Beurskens}, \citenamefont
  {Biedermann}, \citenamefont {Bosch}, \citenamefont {Bozhenkov}, \citenamefont
  {Brakel}, \citenamefont {Dinklage}, \citenamefont {Feng}, \citenamefont
  {Fuchert}, \citenamefont {Geiger}, \citenamefont {Grulke}, \citenamefont
  {Helander}, \citenamefont {Hirsch}, \citenamefont {H{\"o}fel}, \citenamefont
  {Jakubowski}, \citenamefont {Knauer}, \citenamefont {Kocsis}, \citenamefont
  {K{\"o}nig}, \citenamefont {Kornejew}, \citenamefont {Krämer-Flecken},
  \citenamefont {Krychowiak}, \citenamefont {Landreman}, \citenamefont
  {Langenberg}, \citenamefont {Laqua}, \citenamefont {Lazerson}, \citenamefont
  {Maa{\ss}berg}, \citenamefont {Marsen}, \citenamefont {Marushchenko},
  \citenamefont {Moseev}, \citenamefont {Niemann}, \citenamefont {Pablant},
  \citenamefont {Pasch}, \citenamefont {Rahbarnia}, \citenamefont {Schlisio},
  \citenamefont {Stange}, \citenamefont {Pedersen}, \citenamefont {Svensson},
  \citenamefont {Szepesi}, \citenamefont {Mora}, \citenamefont {Turkin},
  \citenamefont {Wauters}, \citenamefont {Weir}, \citenamefont {Wenzel},
  \citenamefont {Windisch}, \citenamefont {Wurden}, \citenamefont {Zhang},
  \citenamefont {Abramovic}, \citenamefont {Äkäslompolo}, \citenamefont
  {Aleynikov}, \citenamefont {Aleynikova}, \citenamefont {Alzbutas},
  \citenamefont {Anda}, \citenamefont {Andreeva}, \citenamefont {Ascasibar},
  \citenamefont {Assmann}, \citenamefont {Baek}, \citenamefont {Banduch},
  \citenamefont {Barbui}, \citenamefont {Barlak}, \citenamefont {Baumann},
  \citenamefont {Behr}, \citenamefont {Benndorf}, \citenamefont {Bertuch},
  \citenamefont {Biel}, \citenamefont {Birus}, \citenamefont {Blackwell},
  \citenamefont {Blanco}, \citenamefont {Blatzheim}, \citenamefont {Bluhm},
  \citenamefont {B{\"o}ckenhoff}, \citenamefont {Bolgert}, \citenamefont
  {Borchardt}, \citenamefont {Borsuk}, \citenamefont {Boscary}, \citenamefont
  {B{\"o}ttger}, \citenamefont {Brand}, \citenamefont {Brandt}, \citenamefont
  {Bräuer}, \citenamefont {Braune}, \citenamefont {Brezinsek}, \citenamefont
  {Brunner}, \citenamefont {Brünner}, \citenamefont {Burhenn}, \citenamefont
  {Buttensch{\"o}n}, \citenamefont {Bykov}, \citenamefont {Calvo},
  \citenamefont {Cannas}, \citenamefont {Cappa}, \citenamefont {Carls},
  \citenamefont {Carraro}, \citenamefont {Carvalho}, \citenamefont {Castejon},
  \citenamefont {Charl}, \citenamefont {Chernyshev}, \citenamefont {Cianciosa},
  \citenamefont {Citarella}, \citenamefont {Ciupi{\'{n}}ski}, \citenamefont
  {Claps}, \citenamefont {Cole}, \citenamefont {Cole}, \citenamefont
  {Cordella}, \citenamefont {Cseh}, \citenamefont {Czarnecka}, \citenamefont
  {Czermak}, \citenamefont {Czerski}, \citenamefont {Czerwinski}, \citenamefont
  {Czymek}, \citenamefont {da~Molin}, \citenamefont {da~Silva}, \citenamefont
  {Dammertz}, \citenamefont {Danielson}, \citenamefont {de~la Pena},
  \citenamefont {Degenkolbe}, \citenamefont {Denner}, \citenamefont {Dhard},
  \citenamefont {Dostal}, \citenamefont {Drevlak}, \citenamefont {Drewelow},
  \citenamefont {Drews}, \citenamefont {Dudek}, \citenamefont {Dundulis},
  \citenamefont {Durodie}, \citenamefont {van Eeten}, \citenamefont
  {Effenberg}, \citenamefont {Ehrke}, \citenamefont {Endler}, \citenamefont
  {Ennis}, \citenamefont {Erckmann}, \citenamefont {Esteban}, \citenamefont
  {Estrada}, \citenamefont {Fahrenkamp}, \citenamefont {Feist}, \citenamefont
  {Fellinger}, \citenamefont {Fernandes}, \citenamefont {Fietz}, \citenamefont
  {Figacz}, \citenamefont {Fontdecaba}, \citenamefont {Ford}, \citenamefont
  {Fornal}, \citenamefont {Frerichs}, \citenamefont {Freund}, \citenamefont
  {Führer}, \citenamefont {Funaba}, \citenamefont {Galkowski}, \citenamefont
  {Gantenbein}, \citenamefont {Gao}, \citenamefont {Rega{\~{n}}a},
  \citenamefont {Garcia-Munoz}, \citenamefont {Gates}, \citenamefont {Gawlik},
  \citenamefont {Geiger}, \citenamefont {Giannella}, \citenamefont {Gierse},
  \citenamefont {Gogoleva}, \citenamefont {Goncalves}, \citenamefont {Goriaev},
  \citenamefont {Gradic}, \citenamefont {Grahl}, \citenamefont {Green},
  \citenamefont {Grosman}, \citenamefont {Grote}, \citenamefont {Gruca},
  \citenamefont {Guerard}, \citenamefont {Haiduk}, \citenamefont {Han},
  \citenamefont {Harberts}, \citenamefont {Harris}, \citenamefont
  {Hartfu{\ss}}, \citenamefont {Hartmann}, \citenamefont {Hathiramani},
  \citenamefont {Hein}, \citenamefont {Heinemann}, \citenamefont
  {Heitzenroeder}, \citenamefont {Henneberg}, \citenamefont {Hennig},
  \citenamefont {Sanchez}, \citenamefont {Hidalgo}, \citenamefont {H{\"o}lbe},
  \citenamefont {Hollfeld}, \citenamefont {H{\"o}lting}, \citenamefont
  {H{\"o}schen}, \citenamefont {Houry}, \citenamefont {Howard}, \citenamefont
  {Huang}, \citenamefont {Huber}, \citenamefont {Huber}, \citenamefont
  {Hunger}, \citenamefont {Ida}, \citenamefont {Ilkei}, \citenamefont {Illy},
  \citenamefont {Israeli}, \citenamefont {Ivanov}, \citenamefont {Jablonski},
  \citenamefont {Jagielski}, \citenamefont {Jelonnek}, \citenamefont {Jenzsch},
  \citenamefont {Junghans}, \citenamefont {Kacmarczyk}, \citenamefont
  {Kaliatka}, \citenamefont {Kallmeyer}, \citenamefont {Kamionka},
  \citenamefont {Karalevicius}, \citenamefont {Kasahara}, \citenamefont
  {Kasparek}, \citenamefont {Kenmochi}, \citenamefont {Keunecke}, \citenamefont
  {Khilchenko}, \citenamefont {Kinna}, \citenamefont {Kleiber}, \citenamefont
  {Klinger}, \citenamefont {Knaup}, \citenamefont {Kobarg}, \citenamefont
  {K{\"o}chl}, \citenamefont {Kolesnichenko}, \citenamefont {K{\"o}nies},
  \citenamefont {K{\"o}ppen}, \citenamefont {Koshurinov}, \citenamefont
  {Koslowski}, \citenamefont {K{\"o}ster}, \citenamefont {Koziol},
  \citenamefont {Krämer}, \citenamefont {Krampitz}, \citenamefont {Kraszewsk},
  \citenamefont {Krawczyk}, \citenamefont {Kremeyer}, \citenamefont {Krings},
  \citenamefont {Krom}, \citenamefont {Krzesinski}, \citenamefont {Ksiazek},
  \citenamefont {Kubkowska}, \citenamefont {Kühner}, \citenamefont
  {Kurki-Suonio}, \citenamefont {Kwak}, \citenamefont {Lang}, \citenamefont
  {Langish}, \citenamefont {Laqua}, \citenamefont {Laube}, \citenamefont
  {Lechte}, \citenamefont {Lennartz}, \citenamefont {Leonhardt}, \citenamefont
  {Lewerentz}, \citenamefont {Liang}, \citenamefont {Linsmeier}, \citenamefont
  {Liu}, \citenamefont {Lobsien}, \citenamefont {Loesser}, \citenamefont
  {Cisquella}, \citenamefont {Lore}, \citenamefont {Lorenz}, \citenamefont
  {Losert}, \citenamefont {Lubyako}, \citenamefont {Lücke}, \citenamefont
  {Lumsdaine}, \citenamefont {Lutsenko}, \citenamefont {Majano-Brown},
  \citenamefont {Marchuk}, \citenamefont {Mardenfeld}, \citenamefont {Marek},
  \citenamefont {Massidda}, \citenamefont {Masuzaki}, \citenamefont {Maurer},
  \citenamefont {McCarthy}, \citenamefont {McNeely}, \citenamefont {Meier},
  \citenamefont {Mellein}, \citenamefont {Mendelevitch}, \citenamefont
  {Mertens}, \citenamefont {Mikkelsen}, \citenamefont {Mishchenko},
  \citenamefont {Missal}, \citenamefont {Mittelstaedt}, \citenamefont
  {Mizuuchi}, \citenamefont {Mollen}, \citenamefont {Moncada}, \citenamefont
  {M{\"o}nnich}, \citenamefont {Morizaki}, \citenamefont {Munk}, \citenamefont
  {Murakami}, \citenamefont {Musielok}, \citenamefont {N{\'{a}}fr{\'{a}}di},
  \citenamefont {Nagel}, \citenamefont {Naujoks}, \citenamefont {Neilson},
  \citenamefont {Neubauer}, \citenamefont {Neuner}, \citenamefont {Ngo},
  \citenamefont {Nocentini}, \citenamefont {Nührenberg}, \citenamefont
  {Nührenberg}, \citenamefont {Obermayer}, \citenamefont {Offermanns},
  \citenamefont {Ogawa}, \citenamefont {Ongena}, \citenamefont {Oosterbeek},
  \citenamefont {Orozco}, \citenamefont {Otte}, \citenamefont {Rodriguez},
  \citenamefont {Pan}, \citenamefont {Panadero}, \citenamefont {Alvarez},
  \citenamefont {Panin}, \citenamefont {Papenfu{\ss}}, \citenamefont {Paqay},
  \citenamefont {Pavone}, \citenamefont {Pawelec}, \citenamefont {Pelka},
  \citenamefont {Peng}, \citenamefont {Perseo}, \citenamefont {Peterson},
  \citenamefont {Pieper}, \citenamefont {Pilopp}, \citenamefont {Pingel},
  \citenamefont {Pisano}, \citenamefont {Plaum}, \citenamefont {Plunk},
  \citenamefont {Povilaitis}, \citenamefont {Preinhaelter}, \citenamefont
  {Proll}, \citenamefont {Puiatti}, \citenamefont {Sitjes}, \citenamefont
  {Purps}, \citenamefont {Rack}, \citenamefont {R{\'{e}}csei}, \citenamefont
  {Reiman}, \citenamefont {Reiter}, \citenamefont {Remppel}, \citenamefont
  {Renard}, \citenamefont {Riedl}, \citenamefont {Riemann}, \citenamefont
  {Rimkevicius}, \citenamefont {Ri{\ss}e}, \citenamefont {Rodatos},
  \citenamefont {R{\"o}hlinger}, \citenamefont {Rom{\'{e}}}, \citenamefont
  {Rong}, \citenamefont {Roscher}, \citenamefont {Roth}, \citenamefont
  {Rudischhauser}, \citenamefont {Rummel}, \citenamefont {Rummel},
  \citenamefont {Runov}, \citenamefont {Rust}, \citenamefont {Ryc},
  \citenamefont {Ryosuke}, \citenamefont {Sakamoto}, \citenamefont {Samartsev},
  \citenamefont {Sanchez}, \citenamefont {Sano}, \citenamefont {Satake},
  \citenamefont {Satheeswaran}, \citenamefont {Schacht}, \citenamefont
  {Schauer}, \citenamefont {Scherer}, \citenamefont {Schlaich}, \citenamefont
  {Schlüter}, \citenamefont {Schmitt}, \citenamefont {Schmitz}, \citenamefont
  {Schmitz}, \citenamefont {Schmuck}, \citenamefont {Schneider}, \citenamefont
  {Schneider}, \citenamefont {Scholz}, \citenamefont {Scholz}, \citenamefont
  {Schrittwieser}, \citenamefont {Schr{\"o}der}, \citenamefont {Schr{\"o}der},
  \citenamefont {Schroeder}, \citenamefont {Schumacher}, \citenamefont
  {Schweer}, \citenamefont {Shanahan}, \citenamefont {Shikhovtsev},
  \citenamefont {Sibilia}, \citenamefont {Sinha}, \citenamefont {Sipliä},
  \citenamefont {Skodzik}, \citenamefont {Slaby}, \citenamefont {Smith},
  \citenamefont {Spiess}, \citenamefont {Spong}, \citenamefont {Spring},
  \citenamefont {Stadler}, \citenamefont {Standley}, \citenamefont {Stephey},
  \citenamefont {Stoneking}, \citenamefont {Stridde}, \citenamefont {Sulek},
  \citenamefont {Surko}, \citenamefont {Suzuki}, \citenamefont {Szab{\'{o}}},
  \citenamefont {Szabolics}, \citenamefont {Sz{\"o}kefalvi-Nagy}, \citenamefont
  {Tamura}, \citenamefont {Terra}, \citenamefont {Terry}, \citenamefont
  {Thomas}, \citenamefont {Thomsen}, \citenamefont {Thumm}, \citenamefont {von
  Thun}, \citenamefont {Timmermann}, \citenamefont {Titus}, \citenamefont
  {Toi}, \citenamefont {Travere}, \citenamefont {Traverso}, \citenamefont
  {Tretter}, \citenamefont {Tsuchiya}, \citenamefont {Tsujimura}, \citenamefont
  {Tulip{\'{a}}n}, \citenamefont {Turnyanskiy}, \citenamefont {Unterberg},
  \citenamefont {Urban}, \citenamefont {Urbonavicius}, \citenamefont
  {Vakulchyk}, \citenamefont {Valet}, \citenamefont {van Millingen},
  \citenamefont {Vela}, \citenamefont {Velasco}, \citenamefont {Vergote},
  \citenamefont {Vervier}, \citenamefont {Vianello}, \citenamefont {Viebke},
  \citenamefont {Vilbrandt}, \citenamefont {Vork{\"o}rper}, \citenamefont
  {Wadle}, \citenamefont {Wagner}, \citenamefont {Wang}, \citenamefont {Wang},
  \citenamefont {Warmer}, \citenamefont {Wegener}, \citenamefont {Weggen},
  \citenamefont {Wei}, \citenamefont {Wendorf}, \citenamefont {Werner},
  \citenamefont {Wiegel}, \citenamefont {Wilde}, \citenamefont {Winkler},
  \citenamefont {Winters}, \citenamefont {Wolf}, \citenamefont {Wolowski},
  \citenamefont {Wright}, \citenamefont {Xanthopoulos}, \citenamefont {Yamada},
  \citenamefont {Yamada}, \citenamefont {Yasuhara}, \citenamefont {Yokoyama},
  \citenamefont {Zajac}, \citenamefont {Zarnstorff}, \citenamefont {Zeitler},
  \citenamefont {Zhang}, \citenamefont {Zhu}, \citenamefont {Zilker},
  \citenamefont {Zimbal}, \citenamefont {Zocco}, \citenamefont {Zoletnik},\
  and\ \citenamefont {Zuin}}]{Wolf_nf_2017}%
  \BibitemOpen
  \bibfield  {author} {\bibinfo {author} {\bibfnamefont {R.}~\bibnamefont
  {Wolf}}, \bibinfo {author} {\bibfnamefont {A.}~\bibnamefont {Ali}}, \bibinfo
  {author} {\bibfnamefont {A.}~\bibnamefont {Alonso}}, \bibinfo {author}
  {\bibfnamefont {J.}~\bibnamefont {Baldzuhn}}, \bibinfo {author}
  {\bibfnamefont {C.}~\bibnamefont {Beidler}}, \bibinfo {author} {\bibfnamefont
  {M.}~\bibnamefont {Beurskens}}, \bibinfo {author} {\bibfnamefont
  {C.}~\bibnamefont {Biedermann}}, \bibinfo {author} {\bibfnamefont {H.-S.}\
  \bibnamefont {Bosch}}, \bibinfo {author} {\bibfnamefont {S.}~\bibnamefont
  {Bozhenkov}}, \bibinfo {author} {\bibfnamefont {R.}~\bibnamefont {Brakel}},
  \bibinfo {author} {\bibfnamefont {A.}~\bibnamefont {Dinklage}}, \bibinfo
  {author} {\bibfnamefont {Y.}~\bibnamefont {Feng}}, \bibinfo {author}
  {\bibfnamefont {G.}~\bibnamefont {Fuchert}}, \bibinfo {author} {\bibfnamefont
  {J.}~\bibnamefont {Geiger}}, \bibinfo {author} {\bibfnamefont
  {O.}~\bibnamefont {Grulke}}, \bibinfo {author} {\bibfnamefont
  {P.}~\bibnamefont {Helander}}, \bibinfo {author} {\bibfnamefont
  {M.}~\bibnamefont {Hirsch}}, \bibinfo {author} {\bibfnamefont
  {U.}~\bibnamefont {H{\"o}fel}}, \bibinfo {author} {\bibfnamefont
  {M.}~\bibnamefont {Jakubowski}}, \bibinfo {author} {\bibfnamefont
  {J.}~\bibnamefont {Knauer}}, \bibinfo {author} {\bibfnamefont
  {G.}~\bibnamefont {Kocsis}}, \bibinfo {author} {\bibfnamefont
  {R.}~\bibnamefont {K{\"o}nig}}, \bibinfo {author} {\bibfnamefont
  {P.}~\bibnamefont {Kornejew}}, \bibinfo {author} {\bibfnamefont
  {A.}~\bibnamefont {Krämer-Flecken}}, \bibinfo {author} {\bibfnamefont
  {M.}~\bibnamefont {Krychowiak}}, \bibinfo {author} {\bibfnamefont
  {M.}~\bibnamefont {Landreman}}, \bibinfo {author} {\bibfnamefont
  {A.}~\bibnamefont {Langenberg}}, \bibinfo {author} {\bibfnamefont
  {H.}~\bibnamefont {Laqua}}, \bibinfo {author} {\bibfnamefont
  {S.}~\bibnamefont {Lazerson}}, \bibinfo {author} {\bibfnamefont
  {H.}~\bibnamefont {Maa{\ss}berg}}, \bibinfo {author} {\bibfnamefont
  {S.}~\bibnamefont {Marsen}}, \bibinfo {author} {\bibfnamefont
  {M.}~\bibnamefont {Marushchenko}}, \bibinfo {author} {\bibfnamefont
  {D.}~\bibnamefont {Moseev}}, \bibinfo {author} {\bibfnamefont
  {H.}~\bibnamefont {Niemann}}, \bibinfo {author} {\bibfnamefont
  {N.}~\bibnamefont {Pablant}}, \bibinfo {author} {\bibfnamefont
  {E.}~\bibnamefont {Pasch}}, \bibinfo {author} {\bibfnamefont
  {K.}~\bibnamefont {Rahbarnia}}, \bibinfo {author} {\bibfnamefont
  {G.}~\bibnamefont {Schlisio}}, \bibinfo {author} {\bibfnamefont
  {T.}~\bibnamefont {Stange}}, \bibinfo {author} {\bibfnamefont {T.~S.}\
  \bibnamefont {Pedersen}}, \bibinfo {author} {\bibfnamefont {J.}~\bibnamefont
  {Svensson}}, \bibinfo {author} {\bibfnamefont {T.}~\bibnamefont {Szepesi}},
  \bibinfo {author} {\bibfnamefont {H.~T.}\ \bibnamefont {Mora}}, \bibinfo
  {author} {\bibfnamefont {Y.}~\bibnamefont {Turkin}}, \bibinfo {author}
  {\bibfnamefont {T.}~\bibnamefont {Wauters}}, \bibinfo {author} {\bibfnamefont
  {G.}~\bibnamefont {Weir}}, \bibinfo {author} {\bibfnamefont {U.}~\bibnamefont
  {Wenzel}}, \bibinfo {author} {\bibfnamefont {T.}~\bibnamefont {Windisch}},
  \bibinfo {author} {\bibfnamefont {G.}~\bibnamefont {Wurden}}, \bibinfo
  {author} {\bibfnamefont {D.}~\bibnamefont {Zhang}}, \bibinfo {author}
  {\bibfnamefont {I.}~\bibnamefont {Abramovic}}, \bibinfo {author}
  {\bibfnamefont {S.}~\bibnamefont {Äkäslompolo}}, \bibinfo {author}
  {\bibfnamefont {P.}~\bibnamefont {Aleynikov}}, \bibinfo {author}
  {\bibfnamefont {K.}~\bibnamefont {Aleynikova}}, \bibinfo {author}
  {\bibfnamefont {R.}~\bibnamefont {Alzbutas}}, \bibinfo {author}
  {\bibfnamefont {G.}~\bibnamefont {Anda}}, \bibinfo {author} {\bibfnamefont
  {T.}~\bibnamefont {Andreeva}}, \bibinfo {author} {\bibfnamefont
  {E.}~\bibnamefont {Ascasibar}}, \bibinfo {author} {\bibfnamefont
  {J.}~\bibnamefont {Assmann}}, \bibinfo {author} {\bibfnamefont {S.-G.}\
  \bibnamefont {Baek}}, \bibinfo {author} {\bibfnamefont {M.}~\bibnamefont
  {Banduch}}, \bibinfo {author} {\bibfnamefont {T.}~\bibnamefont {Barbui}},
  \bibinfo {author} {\bibfnamefont {M.}~\bibnamefont {Barlak}}, \bibinfo
  {author} {\bibfnamefont {K.}~\bibnamefont {Baumann}}, \bibinfo {author}
  {\bibfnamefont {W.}~\bibnamefont {Behr}}, \bibinfo {author} {\bibfnamefont
  {A.}~\bibnamefont {Benndorf}}, \bibinfo {author} {\bibfnamefont
  {O.}~\bibnamefont {Bertuch}}, \bibinfo {author} {\bibfnamefont
  {W.}~\bibnamefont {Biel}}, \bibinfo {author} {\bibfnamefont {D.}~\bibnamefont
  {Birus}}, \bibinfo {author} {\bibfnamefont {B.}~\bibnamefont {Blackwell}},
  \bibinfo {author} {\bibfnamefont {E.}~\bibnamefont {Blanco}}, \bibinfo
  {author} {\bibfnamefont {M.}~\bibnamefont {Blatzheim}}, \bibinfo {author}
  {\bibfnamefont {T.}~\bibnamefont {Bluhm}}, \bibinfo {author} {\bibfnamefont
  {D.}~\bibnamefont {B{\"o}ckenhoff}}, \bibinfo {author} {\bibfnamefont
  {P.}~\bibnamefont {Bolgert}}, \bibinfo {author} {\bibfnamefont
  {M.}~\bibnamefont {Borchardt}}, \bibinfo {author} {\bibfnamefont
  {V.}~\bibnamefont {Borsuk}}, \bibinfo {author} {\bibfnamefont
  {J.}~\bibnamefont {Boscary}}, \bibinfo {author} {\bibfnamefont {L.-G.}\
  \bibnamefont {B{\"o}ttger}}, \bibinfo {author} {\bibfnamefont
  {H.}~\bibnamefont {Brand}}, \bibinfo {author} {\bibfnamefont
  {C.}~\bibnamefont {Brandt}}, \bibinfo {author} {\bibfnamefont
  {T.}~\bibnamefont {Bräuer}}, \bibinfo {author} {\bibfnamefont
  {H.}~\bibnamefont {Braune}}, \bibinfo {author} {\bibfnamefont
  {S.}~\bibnamefont {Brezinsek}}, \bibinfo {author} {\bibfnamefont {K.-J.}\
  \bibnamefont {Brunner}}, \bibinfo {author} {\bibfnamefont {B.}~\bibnamefont
  {Brünner}}, \bibinfo {author} {\bibfnamefont {R.}~\bibnamefont {Burhenn}},
  \bibinfo {author} {\bibfnamefont {B.}~\bibnamefont {Buttensch{\"o}n}},
  \bibinfo {author} {\bibfnamefont {V.}~\bibnamefont {Bykov}}, \bibinfo
  {author} {\bibfnamefont {I.}~\bibnamefont {Calvo}}, \bibinfo {author}
  {\bibfnamefont {B.}~\bibnamefont {Cannas}}, \bibinfo {author} {\bibfnamefont
  {A.}~\bibnamefont {Cappa}}, \bibinfo {author} {\bibfnamefont
  {A.}~\bibnamefont {Carls}}, \bibinfo {author} {\bibfnamefont
  {L.}~\bibnamefont {Carraro}}, \bibinfo {author} {\bibfnamefont
  {B.}~\bibnamefont {Carvalho}}, \bibinfo {author} {\bibfnamefont
  {F.}~\bibnamefont {Castejon}}, \bibinfo {author} {\bibfnamefont
  {A.}~\bibnamefont {Charl}}, \bibinfo {author} {\bibfnamefont
  {F.}~\bibnamefont {Chernyshev}}, \bibinfo {author} {\bibfnamefont
  {M.}~\bibnamefont {Cianciosa}}, \bibinfo {author} {\bibfnamefont
  {R.}~\bibnamefont {Citarella}}, \bibinfo {author} {\bibfnamefont
  {{\L}.}~\bibnamefont {Ciupi{\'{n}}ski}}, \bibinfo {author} {\bibfnamefont
  {G.}~\bibnamefont {Claps}}, \bibinfo {author} {\bibfnamefont
  {M.}~\bibnamefont {Cole}}, \bibinfo {author} {\bibfnamefont {M.}~\bibnamefont
  {Cole}}, \bibinfo {author} {\bibfnamefont {F.}~\bibnamefont {Cordella}},
  \bibinfo {author} {\bibfnamefont {G.}~\bibnamefont {Cseh}}, \bibinfo {author}
  {\bibfnamefont {A.}~\bibnamefont {Czarnecka}}, \bibinfo {author}
  {\bibfnamefont {A.}~\bibnamefont {Czermak}}, \bibinfo {author} {\bibfnamefont
  {K.}~\bibnamefont {Czerski}}, \bibinfo {author} {\bibfnamefont
  {M.}~\bibnamefont {Czerwinski}}, \bibinfo {author} {\bibfnamefont
  {G.}~\bibnamefont {Czymek}}, \bibinfo {author} {\bibfnamefont
  {A.}~\bibnamefont {da~Molin}}, \bibinfo {author} {\bibfnamefont
  {A.}~\bibnamefont {da~Silva}}, \bibinfo {author} {\bibfnamefont
  {G.}~\bibnamefont {Dammertz}}, \bibinfo {author} {\bibfnamefont
  {J.}~\bibnamefont {Danielson}}, \bibinfo {author} {\bibfnamefont
  {A.}~\bibnamefont {de~la Pena}}, \bibinfo {author} {\bibfnamefont
  {S.}~\bibnamefont {Degenkolbe}}, \bibinfo {author} {\bibfnamefont
  {P.}~\bibnamefont {Denner}}, \bibinfo {author} {\bibfnamefont
  {D.}~\bibnamefont {Dhard}}, \bibinfo {author} {\bibfnamefont
  {M.}~\bibnamefont {Dostal}}, \bibinfo {author} {\bibfnamefont
  {M.}~\bibnamefont {Drevlak}}, \bibinfo {author} {\bibfnamefont
  {P.}~\bibnamefont {Drewelow}}, \bibinfo {author} {\bibfnamefont
  {P.}~\bibnamefont {Drews}}, \bibinfo {author} {\bibfnamefont
  {A.}~\bibnamefont {Dudek}}, \bibinfo {author} {\bibfnamefont
  {G.}~\bibnamefont {Dundulis}}, \bibinfo {author} {\bibfnamefont
  {F.}~\bibnamefont {Durodie}}, \bibinfo {author} {\bibfnamefont
  {P.}~\bibnamefont {van Eeten}}, \bibinfo {author} {\bibfnamefont
  {F.}~\bibnamefont {Effenberg}}, \bibinfo {author} {\bibfnamefont
  {G.}~\bibnamefont {Ehrke}}, \bibinfo {author} {\bibfnamefont
  {M.}~\bibnamefont {Endler}}, \bibinfo {author} {\bibfnamefont
  {D.}~\bibnamefont {Ennis}}, \bibinfo {author} {\bibfnamefont
  {E.}~\bibnamefont {Erckmann}}, \bibinfo {author} {\bibfnamefont
  {H.}~\bibnamefont {Esteban}}, \bibinfo {author} {\bibfnamefont
  {T.}~\bibnamefont {Estrada}}, \bibinfo {author} {\bibfnamefont
  {N.}~\bibnamefont {Fahrenkamp}}, \bibinfo {author} {\bibfnamefont {J.-H.}\
  \bibnamefont {Feist}}, \bibinfo {author} {\bibfnamefont {J.}~\bibnamefont
  {Fellinger}}, \bibinfo {author} {\bibfnamefont {H.}~\bibnamefont
  {Fernandes}}, \bibinfo {author} {\bibfnamefont {W.}~\bibnamefont {Fietz}},
  \bibinfo {author} {\bibfnamefont {W.}~\bibnamefont {Figacz}}, \bibinfo
  {author} {\bibfnamefont {J.}~\bibnamefont {Fontdecaba}}, \bibinfo {author}
  {\bibfnamefont {O.}~\bibnamefont {Ford}}, \bibinfo {author} {\bibfnamefont
  {T.}~\bibnamefont {Fornal}}, \bibinfo {author} {\bibfnamefont
  {H.}~\bibnamefont {Frerichs}}, \bibinfo {author} {\bibfnamefont
  {A.}~\bibnamefont {Freund}}, \bibinfo {author} {\bibfnamefont
  {M.}~\bibnamefont {Führer}}, \bibinfo {author} {\bibfnamefont
  {T.}~\bibnamefont {Funaba}}, \bibinfo {author} {\bibfnamefont
  {A.}~\bibnamefont {Galkowski}}, \bibinfo {author} {\bibfnamefont
  {G.}~\bibnamefont {Gantenbein}}, \bibinfo {author} {\bibfnamefont
  {Y.}~\bibnamefont {Gao}}, \bibinfo {author} {\bibfnamefont {J.~G.}\
  \bibnamefont {Rega{\~{n}}a}}, \bibinfo {author} {\bibfnamefont
  {M.}~\bibnamefont {Garcia-Munoz}}, \bibinfo {author} {\bibfnamefont
  {D.}~\bibnamefont {Gates}}, \bibinfo {author} {\bibfnamefont
  {G.}~\bibnamefont {Gawlik}}, \bibinfo {author} {\bibfnamefont
  {B.}~\bibnamefont {Geiger}}, \bibinfo {author} {\bibfnamefont
  {V.}~\bibnamefont {Giannella}}, \bibinfo {author} {\bibfnamefont
  {N.}~\bibnamefont {Gierse}}, \bibinfo {author} {\bibfnamefont
  {A.}~\bibnamefont {Gogoleva}}, \bibinfo {author} {\bibfnamefont
  {B.}~\bibnamefont {Goncalves}}, \bibinfo {author} {\bibfnamefont
  {A.}~\bibnamefont {Goriaev}}, \bibinfo {author} {\bibfnamefont
  {D.}~\bibnamefont {Gradic}}, \bibinfo {author} {\bibfnamefont
  {M.}~\bibnamefont {Grahl}}, \bibinfo {author} {\bibfnamefont
  {J.}~\bibnamefont {Green}}, \bibinfo {author} {\bibfnamefont
  {A.}~\bibnamefont {Grosman}}, \bibinfo {author} {\bibfnamefont
  {H.}~\bibnamefont {Grote}}, \bibinfo {author} {\bibfnamefont
  {M.}~\bibnamefont {Gruca}}, \bibinfo {author} {\bibfnamefont
  {C.}~\bibnamefont {Guerard}}, \bibinfo {author} {\bibfnamefont
  {L.}~\bibnamefont {Haiduk}}, \bibinfo {author} {\bibfnamefont
  {X.}~\bibnamefont {Han}}, \bibinfo {author} {\bibfnamefont {F.}~\bibnamefont
  {Harberts}}, \bibinfo {author} {\bibfnamefont {J.}~\bibnamefont {Harris}},
  \bibinfo {author} {\bibfnamefont {H.-J.}\ \bibnamefont {Hartfu{\ss}}},
  \bibinfo {author} {\bibfnamefont {D.}~\bibnamefont {Hartmann}}, \bibinfo
  {author} {\bibfnamefont {D.}~\bibnamefont {Hathiramani}}, \bibinfo {author}
  {\bibfnamefont {B.}~\bibnamefont {Hein}}, \bibinfo {author} {\bibfnamefont
  {B.}~\bibnamefont {Heinemann}}, \bibinfo {author} {\bibfnamefont
  {P.}~\bibnamefont {Heitzenroeder}}, \bibinfo {author} {\bibfnamefont
  {S.}~\bibnamefont {Henneberg}}, \bibinfo {author} {\bibfnamefont
  {C.}~\bibnamefont {Hennig}}, \bibinfo {author} {\bibfnamefont {J.~H.}\
  \bibnamefont {Sanchez}}, \bibinfo {author} {\bibfnamefont {C.}~\bibnamefont
  {Hidalgo}}, \bibinfo {author} {\bibfnamefont {H.}~\bibnamefont {H{\"o}lbe}},
  \bibinfo {author} {\bibfnamefont {K.}~\bibnamefont {Hollfeld}}, \bibinfo
  {author} {\bibfnamefont {A.}~\bibnamefont {H{\"o}lting}}, \bibinfo {author}
  {\bibfnamefont {D.}~\bibnamefont {H{\"o}schen}}, \bibinfo {author}
  {\bibfnamefont {M.}~\bibnamefont {Houry}}, \bibinfo {author} {\bibfnamefont
  {J.}~\bibnamefont {Howard}}, \bibinfo {author} {\bibfnamefont
  {X.}~\bibnamefont {Huang}}, \bibinfo {author} {\bibfnamefont
  {M.}~\bibnamefont {Huber}}, \bibinfo {author} {\bibfnamefont
  {V.}~\bibnamefont {Huber}}, \bibinfo {author} {\bibfnamefont
  {H.}~\bibnamefont {Hunger}}, \bibinfo {author} {\bibfnamefont
  {K.}~\bibnamefont {Ida}}, \bibinfo {author} {\bibfnamefont {T.}~\bibnamefont
  {Ilkei}}, \bibinfo {author} {\bibfnamefont {S.}~\bibnamefont {Illy}},
  \bibinfo {author} {\bibfnamefont {B.}~\bibnamefont {Israeli}}, \bibinfo
  {author} {\bibfnamefont {A.}~\bibnamefont {Ivanov}}, \bibinfo {author}
  {\bibfnamefont {S.}~\bibnamefont {Jablonski}}, \bibinfo {author}
  {\bibfnamefont {J.}~\bibnamefont {Jagielski}}, \bibinfo {author}
  {\bibfnamefont {J.}~\bibnamefont {Jelonnek}}, \bibinfo {author}
  {\bibfnamefont {H.}~\bibnamefont {Jenzsch}}, \bibinfo {author} {\bibfnamefont
  {P.}~\bibnamefont {Junghans}}, \bibinfo {author} {\bibfnamefont
  {J.}~\bibnamefont {Kacmarczyk}}, \bibinfo {author} {\bibfnamefont
  {T.}~\bibnamefont {Kaliatka}}, \bibinfo {author} {\bibfnamefont {J.-P.}\
  \bibnamefont {Kallmeyer}}, \bibinfo {author} {\bibfnamefont {U.}~\bibnamefont
  {Kamionka}}, \bibinfo {author} {\bibfnamefont {R.}~\bibnamefont
  {Karalevicius}}, \bibinfo {author} {\bibfnamefont {H.}~\bibnamefont
  {Kasahara}}, \bibinfo {author} {\bibfnamefont {W.}~\bibnamefont {Kasparek}},
  \bibinfo {author} {\bibfnamefont {N.}~\bibnamefont {Kenmochi}}, \bibinfo
  {author} {\bibfnamefont {M.}~\bibnamefont {Keunecke}}, \bibinfo {author}
  {\bibfnamefont {A.}~\bibnamefont {Khilchenko}}, \bibinfo {author}
  {\bibfnamefont {D.}~\bibnamefont {Kinna}}, \bibinfo {author} {\bibfnamefont
  {R.}~\bibnamefont {Kleiber}}, \bibinfo {author} {\bibfnamefont
  {T.}~\bibnamefont {Klinger}}, \bibinfo {author} {\bibfnamefont
  {M.}~\bibnamefont {Knaup}}, \bibinfo {author} {\bibfnamefont
  {T.}~\bibnamefont {Kobarg}}, \bibinfo {author} {\bibfnamefont
  {F.}~\bibnamefont {K{\"o}chl}}, \bibinfo {author} {\bibfnamefont
  {Y.}~\bibnamefont {Kolesnichenko}}, \bibinfo {author} {\bibfnamefont
  {A.}~\bibnamefont {K{\"o}nies}}, \bibinfo {author} {\bibfnamefont
  {M.}~\bibnamefont {K{\"o}ppen}}, \bibinfo {author} {\bibfnamefont
  {J.}~\bibnamefont {Koshurinov}}, \bibinfo {author} {\bibfnamefont
  {R.}~\bibnamefont {Koslowski}}, \bibinfo {author} {\bibfnamefont
  {F.}~\bibnamefont {K{\"o}ster}}, \bibinfo {author} {\bibfnamefont
  {R.}~\bibnamefont {Koziol}}, \bibinfo {author} {\bibfnamefont
  {M.}~\bibnamefont {Krämer}}, \bibinfo {author} {\bibfnamefont
  {R.}~\bibnamefont {Krampitz}}, \bibinfo {author} {\bibfnamefont
  {P.}~\bibnamefont {Kraszewsk}}, \bibinfo {author} {\bibfnamefont
  {N.}~\bibnamefont {Krawczyk}}, \bibinfo {author} {\bibfnamefont
  {T.}~\bibnamefont {Kremeyer}}, \bibinfo {author} {\bibfnamefont
  {T.}~\bibnamefont {Krings}}, \bibinfo {author} {\bibfnamefont
  {J.}~\bibnamefont {Krom}}, \bibinfo {author} {\bibfnamefont {G.}~\bibnamefont
  {Krzesinski}}, \bibinfo {author} {\bibfnamefont {I.}~\bibnamefont {Ksiazek}},
  \bibinfo {author} {\bibfnamefont {M.}~\bibnamefont {Kubkowska}}, \bibinfo
  {author} {\bibfnamefont {G.}~\bibnamefont {Kühner}}, \bibinfo {author}
  {\bibfnamefont {T.}~\bibnamefont {Kurki-Suonio}}, \bibinfo {author}
  {\bibfnamefont {S.}~\bibnamefont {Kwak}}, \bibinfo {author} {\bibfnamefont
  {R.}~\bibnamefont {Lang}}, \bibinfo {author} {\bibfnamefont {S.}~\bibnamefont
  {Langish}}, \bibinfo {author} {\bibfnamefont {H.}~\bibnamefont {Laqua}},
  \bibinfo {author} {\bibfnamefont {R.}~\bibnamefont {Laube}}, \bibinfo
  {author} {\bibfnamefont {C.}~\bibnamefont {Lechte}}, \bibinfo {author}
  {\bibfnamefont {M.}~\bibnamefont {Lennartz}}, \bibinfo {author}
  {\bibfnamefont {W.}~\bibnamefont {Leonhardt}}, \bibinfo {author}
  {\bibfnamefont {L.}~\bibnamefont {Lewerentz}}, \bibinfo {author}
  {\bibfnamefont {Y.}~\bibnamefont {Liang}}, \bibinfo {author} {\bibfnamefont
  {C.}~\bibnamefont {Linsmeier}}, \bibinfo {author} {\bibfnamefont
  {S.}~\bibnamefont {Liu}}, \bibinfo {author} {\bibfnamefont {J.-F.}\
  \bibnamefont {Lobsien}}, \bibinfo {author} {\bibfnamefont {D.}~\bibnamefont
  {Loesser}}, \bibinfo {author} {\bibfnamefont {J.~L.}\ \bibnamefont
  {Cisquella}}, \bibinfo {author} {\bibfnamefont {J.}~\bibnamefont {Lore}},
  \bibinfo {author} {\bibfnamefont {A.}~\bibnamefont {Lorenz}}, \bibinfo
  {author} {\bibfnamefont {M.}~\bibnamefont {Losert}}, \bibinfo {author}
  {\bibfnamefont {L.}~\bibnamefont {Lubyako}}, \bibinfo {author} {\bibfnamefont
  {A.}~\bibnamefont {Lücke}}, \bibinfo {author} {\bibfnamefont
  {A.}~\bibnamefont {Lumsdaine}}, \bibinfo {author} {\bibfnamefont
  {V.}~\bibnamefont {Lutsenko}}, \bibinfo {author} {\bibfnamefont
  {J.}~\bibnamefont {Majano-Brown}}, \bibinfo {author} {\bibfnamefont
  {O.}~\bibnamefont {Marchuk}}, \bibinfo {author} {\bibfnamefont
  {M.}~\bibnamefont {Mardenfeld}}, \bibinfo {author} {\bibfnamefont
  {P.}~\bibnamefont {Marek}}, \bibinfo {author} {\bibfnamefont
  {S.}~\bibnamefont {Massidda}}, \bibinfo {author} {\bibfnamefont
  {S.}~\bibnamefont {Masuzaki}}, \bibinfo {author} {\bibfnamefont
  {D.}~\bibnamefont {Maurer}}, \bibinfo {author} {\bibfnamefont
  {K.}~\bibnamefont {McCarthy}}, \bibinfo {author} {\bibfnamefont
  {P.}~\bibnamefont {McNeely}}, \bibinfo {author} {\bibfnamefont
  {A.}~\bibnamefont {Meier}}, \bibinfo {author} {\bibfnamefont
  {D.}~\bibnamefont {Mellein}}, \bibinfo {author} {\bibfnamefont
  {B.}~\bibnamefont {Mendelevitch}}, \bibinfo {author} {\bibfnamefont
  {P.}~\bibnamefont {Mertens}}, \bibinfo {author} {\bibfnamefont
  {D.}~\bibnamefont {Mikkelsen}}, \bibinfo {author} {\bibfnamefont
  {O.}~\bibnamefont {Mishchenko}}, \bibinfo {author} {\bibfnamefont
  {B.}~\bibnamefont {Missal}}, \bibinfo {author} {\bibfnamefont
  {J.}~\bibnamefont {Mittelstaedt}}, \bibinfo {author} {\bibfnamefont
  {T.}~\bibnamefont {Mizuuchi}}, \bibinfo {author} {\bibfnamefont
  {A.}~\bibnamefont {Mollen}}, \bibinfo {author} {\bibfnamefont
  {V.}~\bibnamefont {Moncada}}, \bibinfo {author} {\bibfnamefont
  {T.}~\bibnamefont {M{\"o}nnich}}, \bibinfo {author} {\bibfnamefont
  {T.}~\bibnamefont {Morizaki}}, \bibinfo {author} {\bibfnamefont
  {R.}~\bibnamefont {Munk}}, \bibinfo {author} {\bibfnamefont {S.}~\bibnamefont
  {Murakami}}, \bibinfo {author} {\bibfnamefont {F.}~\bibnamefont {Musielok}},
  \bibinfo {author} {\bibfnamefont {G.}~\bibnamefont {N{\'{a}}fr{\'{a}}di}},
  \bibinfo {author} {\bibfnamefont {M.}~\bibnamefont {Nagel}}, \bibinfo
  {author} {\bibfnamefont {D.}~\bibnamefont {Naujoks}}, \bibinfo {author}
  {\bibfnamefont {H.}~\bibnamefont {Neilson}}, \bibinfo {author} {\bibfnamefont
  {O.}~\bibnamefont {Neubauer}}, \bibinfo {author} {\bibfnamefont
  {U.}~\bibnamefont {Neuner}}, \bibinfo {author} {\bibfnamefont
  {T.}~\bibnamefont {Ngo}}, \bibinfo {author} {\bibfnamefont {R.}~\bibnamefont
  {Nocentini}}, \bibinfo {author} {\bibfnamefont {C.}~\bibnamefont
  {Nührenberg}}, \bibinfo {author} {\bibfnamefont {J.}~\bibnamefont
  {Nührenberg}}, \bibinfo {author} {\bibfnamefont {S.}~\bibnamefont
  {Obermayer}}, \bibinfo {author} {\bibfnamefont {G.}~\bibnamefont
  {Offermanns}}, \bibinfo {author} {\bibfnamefont {K.}~\bibnamefont {Ogawa}},
  \bibinfo {author} {\bibfnamefont {J.}~\bibnamefont {Ongena}}, \bibinfo
  {author} {\bibfnamefont {J.}~\bibnamefont {Oosterbeek}}, \bibinfo {author}
  {\bibfnamefont {G.}~\bibnamefont {Orozco}}, \bibinfo {author} {\bibfnamefont
  {M.}~\bibnamefont {Otte}}, \bibinfo {author} {\bibfnamefont {L.~P.}\
  \bibnamefont {Rodriguez}}, \bibinfo {author} {\bibfnamefont {W.}~\bibnamefont
  {Pan}}, \bibinfo {author} {\bibfnamefont {N.}~\bibnamefont {Panadero}},
  \bibinfo {author} {\bibfnamefont {N.~P.}\ \bibnamefont {Alvarez}}, \bibinfo
  {author} {\bibfnamefont {A.}~\bibnamefont {Panin}}, \bibinfo {author}
  {\bibfnamefont {D.}~\bibnamefont {Papenfu{\ss}}}, \bibinfo {author}
  {\bibfnamefont {S.}~\bibnamefont {Paqay}}, \bibinfo {author} {\bibfnamefont
  {A.}~\bibnamefont {Pavone}}, \bibinfo {author} {\bibfnamefont
  {E.}~\bibnamefont {Pawelec}}, \bibinfo {author} {\bibfnamefont
  {G.}~\bibnamefont {Pelka}}, \bibinfo {author} {\bibfnamefont
  {X.}~\bibnamefont {Peng}}, \bibinfo {author} {\bibfnamefont {V.}~\bibnamefont
  {Perseo}}, \bibinfo {author} {\bibfnamefont {B.}~\bibnamefont {Peterson}},
  \bibinfo {author} {\bibfnamefont {A.}~\bibnamefont {Pieper}}, \bibinfo
  {author} {\bibfnamefont {D.}~\bibnamefont {Pilopp}}, \bibinfo {author}
  {\bibfnamefont {S.}~\bibnamefont {Pingel}}, \bibinfo {author} {\bibfnamefont
  {F.}~\bibnamefont {Pisano}}, \bibinfo {author} {\bibfnamefont
  {B.}~\bibnamefont {Plaum}}, \bibinfo {author} {\bibfnamefont
  {G.}~\bibnamefont {Plunk}}, \bibinfo {author} {\bibfnamefont
  {M.}~\bibnamefont {Povilaitis}}, \bibinfo {author} {\bibfnamefont
  {J.}~\bibnamefont {Preinhaelter}}, \bibinfo {author} {\bibfnamefont
  {J.}~\bibnamefont {Proll}}, \bibinfo {author} {\bibfnamefont {M.-E.}\
  \bibnamefont {Puiatti}}, \bibinfo {author} {\bibfnamefont {A.~P.}\
  \bibnamefont {Sitjes}}, \bibinfo {author} {\bibfnamefont {F.}~\bibnamefont
  {Purps}}, \bibinfo {author} {\bibfnamefont {M.}~\bibnamefont {Rack}},
  \bibinfo {author} {\bibfnamefont {S.}~\bibnamefont {R{\'{e}}csei}}, \bibinfo
  {author} {\bibfnamefont {A.}~\bibnamefont {Reiman}}, \bibinfo {author}
  {\bibfnamefont {D.}~\bibnamefont {Reiter}}, \bibinfo {author} {\bibfnamefont
  {F.}~\bibnamefont {Remppel}}, \bibinfo {author} {\bibfnamefont
  {S.}~\bibnamefont {Renard}}, \bibinfo {author} {\bibfnamefont
  {R.}~\bibnamefont {Riedl}}, \bibinfo {author} {\bibfnamefont
  {J.}~\bibnamefont {Riemann}}, \bibinfo {author} {\bibfnamefont
  {S.}~\bibnamefont {Rimkevicius}}, \bibinfo {author} {\bibfnamefont
  {K.}~\bibnamefont {Ri{\ss}e}}, \bibinfo {author} {\bibfnamefont
  {A.}~\bibnamefont {Rodatos}}, \bibinfo {author} {\bibfnamefont
  {H.}~\bibnamefont {R{\"o}hlinger}}, \bibinfo {author} {\bibfnamefont
  {M.}~\bibnamefont {Rom{\'{e}}}}, \bibinfo {author} {\bibfnamefont
  {P.}~\bibnamefont {Rong}}, \bibinfo {author} {\bibfnamefont {H.-J.}\
  \bibnamefont {Roscher}}, \bibinfo {author} {\bibfnamefont {B.}~\bibnamefont
  {Roth}}, \bibinfo {author} {\bibfnamefont {L.}~\bibnamefont {Rudischhauser}},
  \bibinfo {author} {\bibfnamefont {K.}~\bibnamefont {Rummel}}, \bibinfo
  {author} {\bibfnamefont {T.}~\bibnamefont {Rummel}}, \bibinfo {author}
  {\bibfnamefont {A.}~\bibnamefont {Runov}}, \bibinfo {author} {\bibfnamefont
  {N.}~\bibnamefont {Rust}}, \bibinfo {author} {\bibfnamefont {L.}~\bibnamefont
  {Ryc}}, \bibinfo {author} {\bibfnamefont {S.}~\bibnamefont {Ryosuke}},
  \bibinfo {author} {\bibfnamefont {R.}~\bibnamefont {Sakamoto}}, \bibinfo
  {author} {\bibfnamefont {A.}~\bibnamefont {Samartsev}}, \bibinfo {author}
  {\bibfnamefont {M.}~\bibnamefont {Sanchez}}, \bibinfo {author} {\bibfnamefont
  {F.}~\bibnamefont {Sano}}, \bibinfo {author} {\bibfnamefont {S.}~\bibnamefont
  {Satake}}, \bibinfo {author} {\bibfnamefont {G.}~\bibnamefont
  {Satheeswaran}}, \bibinfo {author} {\bibfnamefont {J.}~\bibnamefont
  {Schacht}}, \bibinfo {author} {\bibfnamefont {F.}~\bibnamefont {Schauer}},
  \bibinfo {author} {\bibfnamefont {T.}~\bibnamefont {Scherer}}, \bibinfo
  {author} {\bibfnamefont {A.}~\bibnamefont {Schlaich}}, \bibinfo {author}
  {\bibfnamefont {K.-H.}\ \bibnamefont {Schlüter}}, \bibinfo {author}
  {\bibfnamefont {J.}~\bibnamefont {Schmitt}}, \bibinfo {author} {\bibfnamefont
  {H.}~\bibnamefont {Schmitz}}, \bibinfo {author} {\bibfnamefont
  {O.}~\bibnamefont {Schmitz}}, \bibinfo {author} {\bibfnamefont
  {S.}~\bibnamefont {Schmuck}}, \bibinfo {author} {\bibfnamefont
  {M.}~\bibnamefont {Schneider}}, \bibinfo {author} {\bibfnamefont
  {W.}~\bibnamefont {Schneider}}, \bibinfo {author} {\bibfnamefont
  {M.}~\bibnamefont {Scholz}}, \bibinfo {author} {\bibfnamefont
  {P.}~\bibnamefont {Scholz}}, \bibinfo {author} {\bibfnamefont
  {R.}~\bibnamefont {Schrittwieser}}, \bibinfo {author} {\bibfnamefont
  {M.}~\bibnamefont {Schr{\"o}der}}, \bibinfo {author} {\bibfnamefont
  {T.}~\bibnamefont {Schr{\"o}der}}, \bibinfo {author} {\bibfnamefont
  {R.}~\bibnamefont {Schroeder}}, \bibinfo {author} {\bibfnamefont
  {H.}~\bibnamefont {Schumacher}}, \bibinfo {author} {\bibfnamefont
  {B.}~\bibnamefont {Schweer}}, \bibinfo {author} {\bibfnamefont
  {B.}~\bibnamefont {Shanahan}}, \bibinfo {author} {\bibfnamefont
  {I.}~\bibnamefont {Shikhovtsev}}, \bibinfo {author} {\bibfnamefont
  {M.}~\bibnamefont {Sibilia}}, \bibinfo {author} {\bibfnamefont
  {P.}~\bibnamefont {Sinha}}, \bibinfo {author} {\bibfnamefont
  {S.}~\bibnamefont {Sipliä}}, \bibinfo {author} {\bibfnamefont
  {J.}~\bibnamefont {Skodzik}}, \bibinfo {author} {\bibfnamefont
  {C.}~\bibnamefont {Slaby}}, \bibinfo {author} {\bibfnamefont
  {H.}~\bibnamefont {Smith}}, \bibinfo {author} {\bibfnamefont
  {W.}~\bibnamefont {Spiess}}, \bibinfo {author} {\bibfnamefont
  {D.}~\bibnamefont {Spong}}, \bibinfo {author} {\bibfnamefont
  {A.}~\bibnamefont {Spring}}, \bibinfo {author} {\bibfnamefont
  {R.}~\bibnamefont {Stadler}}, \bibinfo {author} {\bibfnamefont
  {B.}~\bibnamefont {Standley}}, \bibinfo {author} {\bibfnamefont
  {L.}~\bibnamefont {Stephey}}, \bibinfo {author} {\bibfnamefont
  {M.}~\bibnamefont {Stoneking}}, \bibinfo {author} {\bibfnamefont
  {U.}~\bibnamefont {Stridde}}, \bibinfo {author} {\bibfnamefont
  {Z.}~\bibnamefont {Sulek}}, \bibinfo {author} {\bibfnamefont
  {C.}~\bibnamefont {Surko}}, \bibinfo {author} {\bibfnamefont
  {Y.}~\bibnamefont {Suzuki}}, \bibinfo {author} {\bibfnamefont
  {V.}~\bibnamefont {Szab{\'{o}}}}, \bibinfo {author} {\bibfnamefont
  {T.}~\bibnamefont {Szabolics}}, \bibinfo {author} {\bibfnamefont
  {Z.}~\bibnamefont {Sz{\"o}kefalvi-Nagy}}, \bibinfo {author} {\bibfnamefont
  {N.}~\bibnamefont {Tamura}}, \bibinfo {author} {\bibfnamefont
  {A.}~\bibnamefont {Terra}}, \bibinfo {author} {\bibfnamefont
  {J.}~\bibnamefont {Terry}}, \bibinfo {author} {\bibfnamefont
  {J.}~\bibnamefont {Thomas}}, \bibinfo {author} {\bibfnamefont
  {H.}~\bibnamefont {Thomsen}}, \bibinfo {author} {\bibfnamefont
  {M.}~\bibnamefont {Thumm}}, \bibinfo {author} {\bibfnamefont
  {C.}~\bibnamefont {von Thun}}, \bibinfo {author} {\bibfnamefont
  {D.}~\bibnamefont {Timmermann}}, \bibinfo {author} {\bibfnamefont
  {P.}~\bibnamefont {Titus}}, \bibinfo {author} {\bibfnamefont
  {K.}~\bibnamefont {Toi}}, \bibinfo {author} {\bibfnamefont {J.}~\bibnamefont
  {Travere}}, \bibinfo {author} {\bibfnamefont {P.}~\bibnamefont {Traverso}},
  \bibinfo {author} {\bibfnamefont {J.}~\bibnamefont {Tretter}}, \bibinfo
  {author} {\bibfnamefont {H.}~\bibnamefont {Tsuchiya}}, \bibinfo {author}
  {\bibfnamefont {T.}~\bibnamefont {Tsujimura}}, \bibinfo {author}
  {\bibfnamefont {S.}~\bibnamefont {Tulip{\'{a}}n}}, \bibinfo {author}
  {\bibfnamefont {M.}~\bibnamefont {Turnyanskiy}}, \bibinfo {author}
  {\bibfnamefont {B.}~\bibnamefont {Unterberg}}, \bibinfo {author}
  {\bibfnamefont {J.}~\bibnamefont {Urban}}, \bibinfo {author} {\bibfnamefont
  {E.}~\bibnamefont {Urbonavicius}}, \bibinfo {author} {\bibfnamefont
  {I.}~\bibnamefont {Vakulchyk}}, \bibinfo {author} {\bibfnamefont
  {S.}~\bibnamefont {Valet}}, \bibinfo {author} {\bibfnamefont
  {B.}~\bibnamefont {van Millingen}}, \bibinfo {author} {\bibfnamefont
  {L.}~\bibnamefont {Vela}}, \bibinfo {author} {\bibfnamefont {J.-L.}\
  \bibnamefont {Velasco}}, \bibinfo {author} {\bibfnamefont {M.}~\bibnamefont
  {Vergote}}, \bibinfo {author} {\bibfnamefont {M.}~\bibnamefont {Vervier}},
  \bibinfo {author} {\bibfnamefont {N.}~\bibnamefont {Vianello}}, \bibinfo
  {author} {\bibfnamefont {H.}~\bibnamefont {Viebke}}, \bibinfo {author}
  {\bibfnamefont {R.}~\bibnamefont {Vilbrandt}}, \bibinfo {author}
  {\bibfnamefont {A.}~\bibnamefont {Vork{\"o}rper}}, \bibinfo {author}
  {\bibfnamefont {S.}~\bibnamefont {Wadle}}, \bibinfo {author} {\bibfnamefont
  {F.}~\bibnamefont {Wagner}}, \bibinfo {author} {\bibfnamefont
  {E.}~\bibnamefont {Wang}}, \bibinfo {author} {\bibfnamefont {N.}~\bibnamefont
  {Wang}}, \bibinfo {author} {\bibfnamefont {F.}~\bibnamefont {Warmer}},
  \bibinfo {author} {\bibfnamefont {L.}~\bibnamefont {Wegener}}, \bibinfo
  {author} {\bibfnamefont {J.}~\bibnamefont {Weggen}}, \bibinfo {author}
  {\bibfnamefont {Y.}~\bibnamefont {Wei}}, \bibinfo {author} {\bibfnamefont
  {J.}~\bibnamefont {Wendorf}}, \bibinfo {author} {\bibfnamefont
  {A.}~\bibnamefont {Werner}}, \bibinfo {author} {\bibfnamefont
  {B.}~\bibnamefont {Wiegel}}, \bibinfo {author} {\bibfnamefont
  {F.}~\bibnamefont {Wilde}}, \bibinfo {author} {\bibfnamefont
  {E.}~\bibnamefont {Winkler}}, \bibinfo {author} {\bibfnamefont
  {V.}~\bibnamefont {Winters}}, \bibinfo {author} {\bibfnamefont
  {S.}~\bibnamefont {Wolf}}, \bibinfo {author} {\bibfnamefont {J.}~\bibnamefont
  {Wolowski}}, \bibinfo {author} {\bibfnamefont {A.}~\bibnamefont {Wright}},
  \bibinfo {author} {\bibfnamefont {P.}~\bibnamefont {Xanthopoulos}}, \bibinfo
  {author} {\bibfnamefont {H.}~\bibnamefont {Yamada}}, \bibinfo {author}
  {\bibfnamefont {I.}~\bibnamefont {Yamada}}, \bibinfo {author} {\bibfnamefont
  {R.}~\bibnamefont {Yasuhara}}, \bibinfo {author} {\bibfnamefont
  {M.}~\bibnamefont {Yokoyama}}, \bibinfo {author} {\bibfnamefont
  {J.}~\bibnamefont {Zajac}}, \bibinfo {author} {\bibfnamefont
  {M.}~\bibnamefont {Zarnstorff}}, \bibinfo {author} {\bibfnamefont
  {A.}~\bibnamefont {Zeitler}}, \bibinfo {author} {\bibfnamefont
  {H.}~\bibnamefont {Zhang}}, \bibinfo {author} {\bibfnamefont
  {J.}~\bibnamefont {Zhu}}, \bibinfo {author} {\bibfnamefont {M.}~\bibnamefont
  {Zilker}}, \bibinfo {author} {\bibfnamefont {A.}~\bibnamefont {Zimbal}},
  \bibinfo {author} {\bibfnamefont {A.}~\bibnamefont {Zocco}}, \bibinfo
  {author} {\bibfnamefont {S.}~\bibnamefont {Zoletnik}}, \ and\ \bibinfo
  {author} {\bibfnamefont {M.}~\bibnamefont {Zuin}},\ }\href {\doibase
  10.1088/1741-4326/aa770d} {\bibfield  {journal} {\bibinfo  {journal} {Nuclear
  Fusion}\ }\textbf {\bibinfo {volume} {57}},\ \bibinfo {pages} {102020}
  (\bibinfo {year} {2017})}\BibitemShut {NoStop}%
\bibitem [{\citenamefont {Barnes}\ \emph {et~al.}(2019)\citenamefont {Barnes},
  \citenamefont {Parra},\ and\ \citenamefont {Landreman}}]{Barnes_jcp_2019}%
  \BibitemOpen
  \bibfield  {author} {\bibinfo {author} {\bibfnamefont {M.}~\bibnamefont
  {Barnes}}, \bibinfo {author} {\bibfnamefont {F.}~\bibnamefont {Parra}}, \
  and\ \bibinfo {author} {\bibfnamefont {M.}~\bibnamefont {Landreman}},\ }\href
  {\doibase 10.1016/j.jcp.2019.01.025} {\bibfield  {journal} {\bibinfo
  {journal} {Journal of Computational Physics}\ }\textbf {\bibinfo {volume}
  {391}},\ \bibinfo {pages} {365} (\bibinfo {year} {2019})}\BibitemShut
  {NoStop}%
\bibitem [{\citenamefont {Gonz{\'{a}}lez-Jerez}\ \emph
  {et~al.}(2022)\citenamefont {Gonz{\'{a}}lez-Jerez}, \citenamefont
  {Xanthopoulos}, \citenamefont {Garc{\'{\i}}a-Rega{\~{n}}a}, \citenamefont
  {Calvo}, \citenamefont {Alcus{\'{o}}n}, \citenamefont {Navarro},
  \citenamefont {Barnes}, \citenamefont {Parra},\ and\ \citenamefont
  {Geiger}}]{GonzalezJerez_jpp_2022}%
  \BibitemOpen
  \bibfield  {author} {\bibinfo {author} {\bibfnamefont {A.}~\bibnamefont
  {Gonz{\'{a}}lez-Jerez}}, \bibinfo {author} {\bibfnamefont {P.}~\bibnamefont
  {Xanthopoulos}}, \bibinfo {author} {\bibfnamefont {J.}~\bibnamefont
  {Garc{\'{\i}}a-Rega{\~{n}}a}}, \bibinfo {author} {\bibfnamefont
  {I.}~\bibnamefont {Calvo}}, \bibinfo {author} {\bibfnamefont
  {J.}~\bibnamefont {Alcus{\'{o}}n}}, \bibinfo {author} {\bibfnamefont {A.~B.}\
  \bibnamefont {Navarro}}, \bibinfo {author} {\bibfnamefont {M.}~\bibnamefont
  {Barnes}}, \bibinfo {author} {\bibfnamefont {F.}~\bibnamefont {Parra}}, \
  and\ \bibinfo {author} {\bibfnamefont {J.}~\bibnamefont {Geiger}},\ }\href
  {https://doi.org/10.1017/s0022377822000393} {\bibfield  {journal} {\bibinfo
  {journal} {Journal of Plasma Physics}\ }\textbf {\bibinfo {volume} {88}}
  (\bibinfo {year} {2022})}\BibitemShut {NoStop}%
\bibitem [{\citenamefont {Garc{\'{\i}}a-Rega{\~{n}}a}\ \emph
  {et~al.}(2021{\natexlab{a}})\citenamefont {Garc{\'{\i}}a-Rega{\~{n}}a},
  \citenamefont {Barnes}, \citenamefont {Calvo}, \citenamefont {Parra},
  \citenamefont {Alcus{\'{o}}n}, \citenamefont {Davies}, \citenamefont
  {Gonz{\'{a}}lez-Jerez}, \citenamefont {Moll{\'{e}}n}, \citenamefont
  {S{\'{a}}nchez}, \citenamefont {Velasco},\ and\ \citenamefont
  {Zocco}}]{Regana_jpp_2021}%
  \BibitemOpen
  \bibfield  {author} {\bibinfo {author} {\bibfnamefont {J.~M.}\ \bibnamefont
  {Garc{\'{\i}}a-Rega{\~{n}}a}}, \bibinfo {author} {\bibfnamefont
  {M.}~\bibnamefont {Barnes}}, \bibinfo {author} {\bibfnamefont
  {I.}~\bibnamefont {Calvo}}, \bibinfo {author} {\bibfnamefont {F.~I.}\
  \bibnamefont {Parra}}, \bibinfo {author} {\bibfnamefont {J.~A.}\ \bibnamefont
  {Alcus{\'{o}}n}}, \bibinfo {author} {\bibfnamefont {R.}~\bibnamefont
  {Davies}}, \bibinfo {author} {\bibfnamefont {A.}~\bibnamefont
  {Gonz{\'{a}}lez-Jerez}}, \bibinfo {author} {\bibfnamefont {A.}~\bibnamefont
  {Moll{\'{e}}n}}, \bibinfo {author} {\bibfnamefont {E.}~\bibnamefont
  {S{\'{a}}nchez}}, \bibinfo {author} {\bibfnamefont {J.~L.}\ \bibnamefont
  {Velasco}}, \ and\ \bibinfo {author} {\bibfnamefont {A.}~\bibnamefont
  {Zocco}},\ }\href@noop {} {\bibfield  {journal} {\bibinfo  {journal} {Journal
  of Plasma Physics}\ }\textbf {\bibinfo {volume} {87}} (\bibinfo {year}
  {2021}{\natexlab{a}})}\BibitemShut {NoStop}%
\bibitem [{\citenamefont {Garc{\'{\i}}a-Rega{\~{n}}a}\ \emph
  {et~al.}(2021{\natexlab{b}})\citenamefont {Garc{\'{\i}}a-Rega{\~{n}}a},
  \citenamefont {Barnes}, \citenamefont {Calvo}, \citenamefont
  {Gonz{\'{a}}lez-Jerez}, \citenamefont {Thienpondt}, \citenamefont
  {S{\'{a}}nchez}, \citenamefont {Parra},\ and\ \citenamefont
  {St.-Onge}}]{Regana_nf_2021}%
  \BibitemOpen
  \bibfield  {author} {\bibinfo {author} {\bibfnamefont {J.}~\bibnamefont
  {Garc{\'{\i}}a-Rega{\~{n}}a}}, \bibinfo {author} {\bibfnamefont
  {M.}~\bibnamefont {Barnes}}, \bibinfo {author} {\bibfnamefont
  {I.}~\bibnamefont {Calvo}}, \bibinfo {author} {\bibfnamefont
  {A.}~\bibnamefont {Gonz{\'{a}}lez-Jerez}}, \bibinfo {author} {\bibfnamefont
  {H.}~\bibnamefont {Thienpondt}}, \bibinfo {author} {\bibfnamefont
  {E.}~\bibnamefont {S{\'{a}}nchez}}, \bibinfo {author} {\bibfnamefont {F.~I.}\
  \bibnamefont {Parra}}, \ and\ \bibinfo {author} {\bibfnamefont {D.~A.}\
  \bibnamefont {St.-Onge}},\ }\href {\doibase 10.1088/1741-4326/ac1d84}
  {\bibfield  {journal} {\bibinfo  {journal} {Nuclear Fusion}\ }\textbf
  {\bibinfo {volume} {61}},\ \bibinfo {pages} {116019} (\bibinfo {year}
  {2021}{\natexlab{b}})}\BibitemShut {NoStop}%
\bibitem [{\citenamefont {Martin}\ \emph {et~al.}(2018)\citenamefont {Martin},
  \citenamefont {Landreman}, \citenamefont {Xanthopoulos}, \citenamefont
  {Mandell},\ and\ \citenamefont {Dorland}}]{Martin_ppcf_2018}%
  \BibitemOpen
  \bibfield  {author} {\bibinfo {author} {\bibfnamefont {M.~F.}\ \bibnamefont
  {Martin}}, \bibinfo {author} {\bibfnamefont {M.}~\bibnamefont {Landreman}},
  \bibinfo {author} {\bibfnamefont {P.}~\bibnamefont {Xanthopoulos}}, \bibinfo
  {author} {\bibfnamefont {N.~R.}\ \bibnamefont {Mandell}}, \ and\ \bibinfo
  {author} {\bibfnamefont {W.}~\bibnamefont {Dorland}},\ }\href {\doibase
  10.1088/1361-6587/aad38a} {\bibfield  {journal} {\bibinfo  {journal} {Plasma
  Physics and Controlled Fusion}\ }\textbf {\bibinfo {volume} {60}},\ \bibinfo
  {pages} {095008} (\bibinfo {year} {2018})}\BibitemShut {NoStop}%
\bibitem [{\citenamefont {{H. Thienpondt et al.}}()}]{Thienpondt_ISHW_2022}%
  \BibitemOpen
  \bibfield  {author} {\bibinfo {author} {\bibnamefont {{H. Thienpondt et
  al.}}},\ }\href@noop {} {\emph {\bibinfo {title} {{23rd International
  Stellarator-Heliotron Workshop, Warsaw, Poland, June 20--24,
  2022}}}}\BibitemShut {NoStop}%
\bibitem [{\citenamefont {Ishizawa}\ \emph {et~al.}(2017)\citenamefont
  {Ishizawa}, \citenamefont {Kishimoto}, \citenamefont {Watanabe},
  \citenamefont {Sugama}, \citenamefont {Tanaka}, \citenamefont {Satake},
  \citenamefont {Kobayashi}, \citenamefont {Nagasaki},\ and\ \citenamefont
  {Nakamura}}]{Ishizawa_nf_2017}%
  \BibitemOpen
  \bibfield  {author} {\bibinfo {author} {\bibfnamefont {A.}~\bibnamefont
  {Ishizawa}}, \bibinfo {author} {\bibfnamefont {Y.}~\bibnamefont {Kishimoto}},
  \bibinfo {author} {\bibfnamefont {T.-H.}\ \bibnamefont {Watanabe}}, \bibinfo
  {author} {\bibfnamefont {H.}~\bibnamefont {Sugama}}, \bibinfo {author}
  {\bibfnamefont {K.}~\bibnamefont {Tanaka}}, \bibinfo {author} {\bibfnamefont
  {S.}~\bibnamefont {Satake}}, \bibinfo {author} {\bibfnamefont
  {S.}~\bibnamefont {Kobayashi}}, \bibinfo {author} {\bibfnamefont
  {K.}~\bibnamefont {Nagasaki}}, \ and\ \bibinfo {author} {\bibfnamefont
  {Y.}~\bibnamefont {Nakamura}},\ }\href {\doibase 10.1088/1741-4326/aa6603}
  {\bibfield  {journal} {\bibinfo  {journal} {Nuclear Fusion}\ }\textbf
  {\bibinfo {volume} {57}},\ \bibinfo {pages} {066010} (\bibinfo {year}
  {2017})}\BibitemShut {NoStop}%
\bibitem [{\citenamefont {Nunami}\ \emph {et~al.}(2020)\citenamefont {Nunami},
  \citenamefont {Nakata}, \citenamefont {Toda},\ and\ \citenamefont
  {Sugama}}]{Nunami_pop_2020}%
  \BibitemOpen
  \bibfield  {author} {\bibinfo {author} {\bibfnamefont {M.}~\bibnamefont
  {Nunami}}, \bibinfo {author} {\bibfnamefont {M.}~\bibnamefont {Nakata}},
  \bibinfo {author} {\bibfnamefont {S.}~\bibnamefont {Toda}}, \ and\ \bibinfo
  {author} {\bibfnamefont {H.}~\bibnamefont {Sugama}},\ }\href {\doibase
  10.1063/1.5142405} {\bibfield  {journal} {\bibinfo  {journal} {Physics of
  Plasmas}\ }\textbf {\bibinfo {volume} {27}},\ \bibinfo {pages} {052501}
  (\bibinfo {year} {2020})}\BibitemShut {NoStop}%
\bibitem [{\citenamefont {Kremeyer}\ \emph {et~al.}(2022)\citenamefont
  {Kremeyer}, \citenamefont {K{\"o}nig}, \citenamefont {Brezinsek},
  \citenamefont {Schmitz}, \citenamefont {Feng}, \citenamefont {Winters},
  \citenamefont {Rudischhauser}, \citenamefont {Buttensch{\"o}n}, \citenamefont
  {Brunner}, \citenamefont {Drewelow}, \citenamefont {Flom}, \citenamefont
  {Fuchert}, \citenamefont {Gao}, \citenamefont {Geiger}, \citenamefont
  {Jakubowski}, \citenamefont {Killer}, \citenamefont {Knauer}, \citenamefont
  {Krychowiak}, \citenamefont {Lazerson}, \citenamefont {Reimold},
  \citenamefont {Schlisio}, \citenamefont {Viebke},\ and\ \citenamefont {the
  W7-X~Team}}]{Kremeyer_nf_2022}%
  \BibitemOpen
  \bibfield  {author} {\bibinfo {author} {\bibfnamefont {T.}~\bibnamefont
  {Kremeyer}}, \bibinfo {author} {\bibfnamefont {R.}~\bibnamefont {K{\"o}nig}},
  \bibinfo {author} {\bibfnamefont {S.}~\bibnamefont {Brezinsek}}, \bibinfo
  {author} {\bibfnamefont {O.}~\bibnamefont {Schmitz}}, \bibinfo {author}
  {\bibfnamefont {Y.}~\bibnamefont {Feng}}, \bibinfo {author} {\bibfnamefont
  {V.}~\bibnamefont {Winters}}, \bibinfo {author} {\bibfnamefont
  {L.}~\bibnamefont {Rudischhauser}}, \bibinfo {author} {\bibfnamefont
  {B.}~\bibnamefont {Buttensch{\"o}n}}, \bibinfo {author} {\bibfnamefont
  {K.}~\bibnamefont {Brunner}}, \bibinfo {author} {\bibfnamefont
  {P.}~\bibnamefont {Drewelow}}, \bibinfo {author} {\bibfnamefont
  {E.}~\bibnamefont {Flom}}, \bibinfo {author} {\bibfnamefont {G.}~\bibnamefont
  {Fuchert}}, \bibinfo {author} {\bibfnamefont {Y.}~\bibnamefont {Gao}},
  \bibinfo {author} {\bibfnamefont {J.}~\bibnamefont {Geiger}}, \bibinfo
  {author} {\bibfnamefont {M.}~\bibnamefont {Jakubowski}}, \bibinfo {author}
  {\bibfnamefont {C.}~\bibnamefont {Killer}}, \bibinfo {author} {\bibfnamefont
  {J.}~\bibnamefont {Knauer}}, \bibinfo {author} {\bibfnamefont
  {M.}~\bibnamefont {Krychowiak}}, \bibinfo {author} {\bibfnamefont
  {S.}~\bibnamefont {Lazerson}}, \bibinfo {author} {\bibfnamefont
  {F.}~\bibnamefont {Reimold}}, \bibinfo {author} {\bibfnamefont
  {G.}~\bibnamefont {Schlisio}}, \bibinfo {author} {\bibfnamefont
  {H.}~\bibnamefont {Viebke}}, \ and\ \bibinfo {author} {\bibnamefont {the
  W7-X~Team}},\ }\href {\doibase 10.1088/1741-4326/ac4acb} {\bibfield
  {journal} {\bibinfo  {journal} {Nuclear Fusion}\ }\textbf {\bibinfo {volume}
  {62}},\ \bibinfo {pages} {036023} (\bibinfo {year} {2022})}\BibitemShut
  {NoStop}%
\bibitem [{\citenamefont {Hazeltine}\ \emph {et~al.}(1992)\citenamefont
  {Hazeltine}, \citenamefont {Calvin}, \citenamefont {Valanju},\ and\
  \citenamefont {Solano}}]{Hazeltine_nf_1992}%
  \BibitemOpen
  \bibfield  {author} {\bibinfo {author} {\bibfnamefont {R.}~\bibnamefont
  {Hazeltine}}, \bibinfo {author} {\bibfnamefont {M.}~\bibnamefont {Calvin}},
  \bibinfo {author} {\bibfnamefont {P.}~\bibnamefont {Valanju}}, \ and\
  \bibinfo {author} {\bibfnamefont {E.}~\bibnamefont {Solano}},\ }\href
  {\doibase 10.1088/0029-5515/32/1/i01} {\bibfield  {journal} {\bibinfo
  {journal} {Nuclear Fusion}\ }\textbf {\bibinfo {volume} {32}},\ \bibinfo
  {pages} {3} (\bibinfo {year} {1992})}\BibitemShut {NoStop}%
\bibitem [{\citenamefont {Hirshman}\ \emph {et~al.}(1986)\citenamefont
  {Hirshman}, \citenamefont {Shaing}, \citenamefont {van Rij}, \citenamefont
  {Beasley},\ and\ \citenamefont {Crume}}]{Hirshman1986}%
  \BibitemOpen
  \bibfield  {author} {\bibinfo {author} {\bibfnamefont {S.~P.}\ \bibnamefont
  {Hirshman}}, \bibinfo {author} {\bibfnamefont {K.~C.}\ \bibnamefont
  {Shaing}}, \bibinfo {author} {\bibfnamefont {W.~I.}\ \bibnamefont {van Rij}},
  \bibinfo {author} {\bibfnamefont {C.~O.}\ \bibnamefont {Beasley}}, \ and\
  \bibinfo {author} {\bibfnamefont {E.~C.}\ \bibnamefont {Crume}},\ }\href
  {\doibase 10.1063/1.865495} {\bibfield  {journal} {\bibinfo  {journal} {The
  Physics of Fluids}\ }\textbf {\bibinfo {volume} {29}},\ \bibinfo {pages}
  {2951} (\bibinfo {year} {1986})}\BibitemShut {NoStop}%
\bibitem [{\citenamefont {Canik}\ \emph
  {et~al.}(2007{\natexlab{a}})\citenamefont {Canik}, \citenamefont {Anderson},
  \citenamefont {Anderson}, \citenamefont {Clark}, \citenamefont {Likin},
  \citenamefont {Talmadge},\ and\ \citenamefont {Zhai}}]{HSX1}%
  \BibitemOpen
  \bibfield  {author} {\bibinfo {author} {\bibfnamefont {J.~M.}\ \bibnamefont
  {Canik}}, \bibinfo {author} {\bibfnamefont {D.~T.}\ \bibnamefont {Anderson}},
  \bibinfo {author} {\bibfnamefont {F.~S.~B.}\ \bibnamefont {Anderson}},
  \bibinfo {author} {\bibfnamefont {C.}~\bibnamefont {Clark}}, \bibinfo
  {author} {\bibfnamefont {K.~M.}\ \bibnamefont {Likin}}, \bibinfo {author}
  {\bibfnamefont {J.~N.}\ \bibnamefont {Talmadge}}, \ and\ \bibinfo {author}
  {\bibfnamefont {K.}~\bibnamefont {Zhai}},\ }\href {\doibase
  10.1063/1.2709862} {\bibfield  {journal} {\bibinfo  {journal} {Physics of
  Plasmas}\ }\textbf {\bibinfo {volume} {14}},\ \bibinfo {pages} {056107}
  (\bibinfo {year} {2007}{\natexlab{a}})},\ \Eprint
  {http://arxiv.org/abs/https://doi.org/10.1063/1.2709862}
  {https://doi.org/10.1063/1.2709862} \BibitemShut {NoStop}%
\bibitem [{\citenamefont {Canik}\ \emph
  {et~al.}(2007{\natexlab{b}})\citenamefont {Canik}, \citenamefont {Anderson},
  \citenamefont {Anderson}, \citenamefont {Likin}, \citenamefont {Talmadge},\
  and\ \citenamefont {Zhai}}]{HSX2}%
  \BibitemOpen
  \bibfield  {author} {\bibinfo {author} {\bibfnamefont {J.~M.}\ \bibnamefont
  {Canik}}, \bibinfo {author} {\bibfnamefont {D.~T.}\ \bibnamefont {Anderson}},
  \bibinfo {author} {\bibfnamefont {F.~S.~B.}\ \bibnamefont {Anderson}},
  \bibinfo {author} {\bibfnamefont {K.~M.}\ \bibnamefont {Likin}}, \bibinfo
  {author} {\bibfnamefont {J.~N.}\ \bibnamefont {Talmadge}}, \ and\ \bibinfo
  {author} {\bibfnamefont {K.}~\bibnamefont {Zhai}},\ }\href {\doibase
  10.1103/PhysRevLett.98.085002} {\bibfield  {journal} {\bibinfo  {journal}
  {Phys. Rev. Lett.}\ }\textbf {\bibinfo {volume} {98}},\ \bibinfo {pages}
  {085002} (\bibinfo {year} {2007}{\natexlab{b}})}\BibitemShut {NoStop}%
\bibitem [{\citenamefont {Grimm}\ \emph {et~al.}(1983)\citenamefont {Grimm},
  \citenamefont {Dewar},\ and\ \citenamefont {Manickam}}]{grimm1983ideal}%
  \BibitemOpen
  \bibfield  {author} {\bibinfo {author} {\bibfnamefont {R.}~\bibnamefont
  {Grimm}}, \bibinfo {author} {\bibfnamefont {R.}~\bibnamefont {Dewar}}, \ and\
  \bibinfo {author} {\bibfnamefont {J.}~\bibnamefont {Manickam}},\ }\href
  {\doibase https://doi.org/10.1016/0021-9991(83)90116-X} {\bibfield  {journal}
  {\bibinfo  {journal} {Journal of Computational Physics}\ }\textbf {\bibinfo
  {volume} {49}},\ \bibinfo {pages} {94} (\bibinfo {year} {1983})}\BibitemShut
  {NoStop}%
\end{thebibliography}%

\end{document}